\documentclass[twocolumn,super,citeautoscript]{revtex4}

\usepackage{graphicx}
\usepackage{amsmath}
\usepackage{amssymb}
\usepackage{color}

\newcommand{\ncite}[1]{{\color{blue}\cite{#1}}}
\newcommand{\rcite}[2]{{\color{blue}{#1}}}

\renewcommand{\section}[1]{\noindent{\bf \large #1}}
\renewcommand{\subsection}[1]{\noindent{\bf #1}}

\begin{document}

\title{A weakly-interacting many-body system of Rydberg polaritons based on electromagnetically induced transparency}

\author{Bongjune Kim$^1$}
\author{Ko-Tang Chen$^1$} 
\author{Shih-Si Hsiao$^1$} 
\author{Sheng-Yang Wang$^1$}
\author{Kai-Bo Li$^1$}
\author{Julius Ruseckas$^2$}
\author{Gediminas Juzeli\={u}nas$^2$}
\author{Teodora Kirova$^3$}
\author{Marcis Auzinsh$^4$}
\author{Ying-Cheng Chen$^{5,7}$}
\author{Yong-Fan Chen$^{6,7}$}
\author{Ite A. Yu$^{1,7,*}$}

\affiliation{
$^{1}$Department of Physics, National Tsing Hua University, Hsinchu 30013, Taiwan \\
$^{2}$Institute of Theoretical Physics and Astronomy, Vilnius University, Saul\.{e}tekio 3, 10257 Vilnius, Lithuania \\
$^{3}$Institute of Atomic Physics and Spectroscopy, University of Latvia, LV-1004 Riga,
Latvia \\
$^{4}$Laser Centre, University of Latvia, LV-1002, Riga, Latvia \\
$^{5}$Institute of Atomic and Molecular Sciences, Academia Sinica, Taipei 10617, Taiwan \\
$^{6}$Department of Physics, National Cheng Kung University, Tainan 70101, Taiwan \\
$^{7}$Center for Quantum Technology, Hsinchu 30013, Taiwan \\
\hspace*{1.5cm}*email: yu@phys.nthu.edu.tw~~~~~~~~
}

\begin{abstract}
\section{Abstract}
The combination of Rydberg atoms and electromagnetically induced transparency (EIT) has been extensively studied in the strong-interaction regime. Here we proposed utilizing an EIT medium with a high optical depth (OD) and a Rydberg state of low principal quantum number to create a many-body system of Rydberg polaritons in the weak-interaction regime. The phase shift and attenuation induced by the dipole-dipole interaction (DDI) were still significant, and can be viewed as the consequences of elastic and inelastic collisions among Rydberg polaritons. We further observed that the width of the transverse momentum distribution of Rydberg polaritons at the exit of the system became notably smaller as compared with that at the entrance. The observation demonstrates the cooling effect in this system. The $\mu$s-long interaction time due to the high-OD EIT medium plus the $\mu$m$^2$-size collision cross section due to the DDI suggests a feasible platform of polariton Bose-Einstein condensation.
\end{abstract}

\maketitle

\newcommand{\FigOne}{
	\begin{figure*}[t]
	\center{\includegraphics[width=115mm]{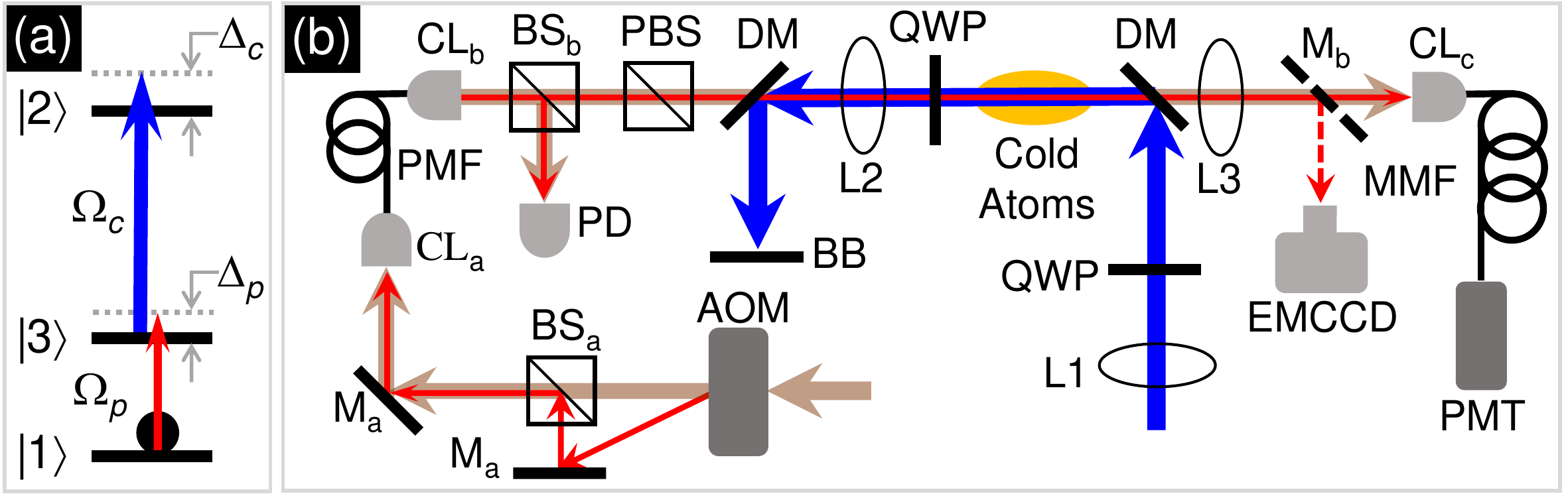}}
	\caption{Transition scheme and experimental setup. (a) Relevant energy levels and laser excitations in the experiment. States $|1\rangle$, $|2\rangle$ and $|3\rangle$ correspond to the ground state $|5S_{1/2}, F=2, m_F=2\rangle$, Rydberg state $|32D_{5/2}, m_J=5/2\rangle$, and excited state $|5P_{3/2}, F=3, m_F=3\rangle$ of $^{87}$Rb atoms. $\Omega_c$ and $\Omega_p$ denote the coupling and probe Rabi frequencies, and $\Delta_c$ and $\Delta_p$ represent their detunings. (b) Experimental setup. AOM: acousto-optic modulator; BS$_{a}$, BS$_{b}$: beam splitters; CL$_{a}$, CL$_{b}$, CL$_{c}$: collimation lenses; PMF: polarization-maintained optical fiber; BB: beam block; M$_{a}$ mirror; M$_{b}$: movable mirror on a flip mount; PD: photo detector (Thorlabs APD110A); PBS: polarizing beam splitter; DM: dichroic mirror; L1, L2, L3: lenses; QWP: quarter-wave plate; EMCCD: electron-multiplying charge-coupled device camera (Andor DL-604M-OEM); MMF: multimode optical fiber; PMT: photomultiplier tube and amplifier (Hamamatsu H6780-20 and C9663). Blue, red, and brown arrowed lines indicate the optical paths of the coupling, probe, and AOM's zeroth-order beams, respectively. 	
}
	\label{fig:EnergyLevel}
	\label{fig:ExpSetup}
	\end{figure*}
}
\newcommand{\FigTwo}{
	\begin{figure*}[t]
	\center{\includegraphics[width=105mm]{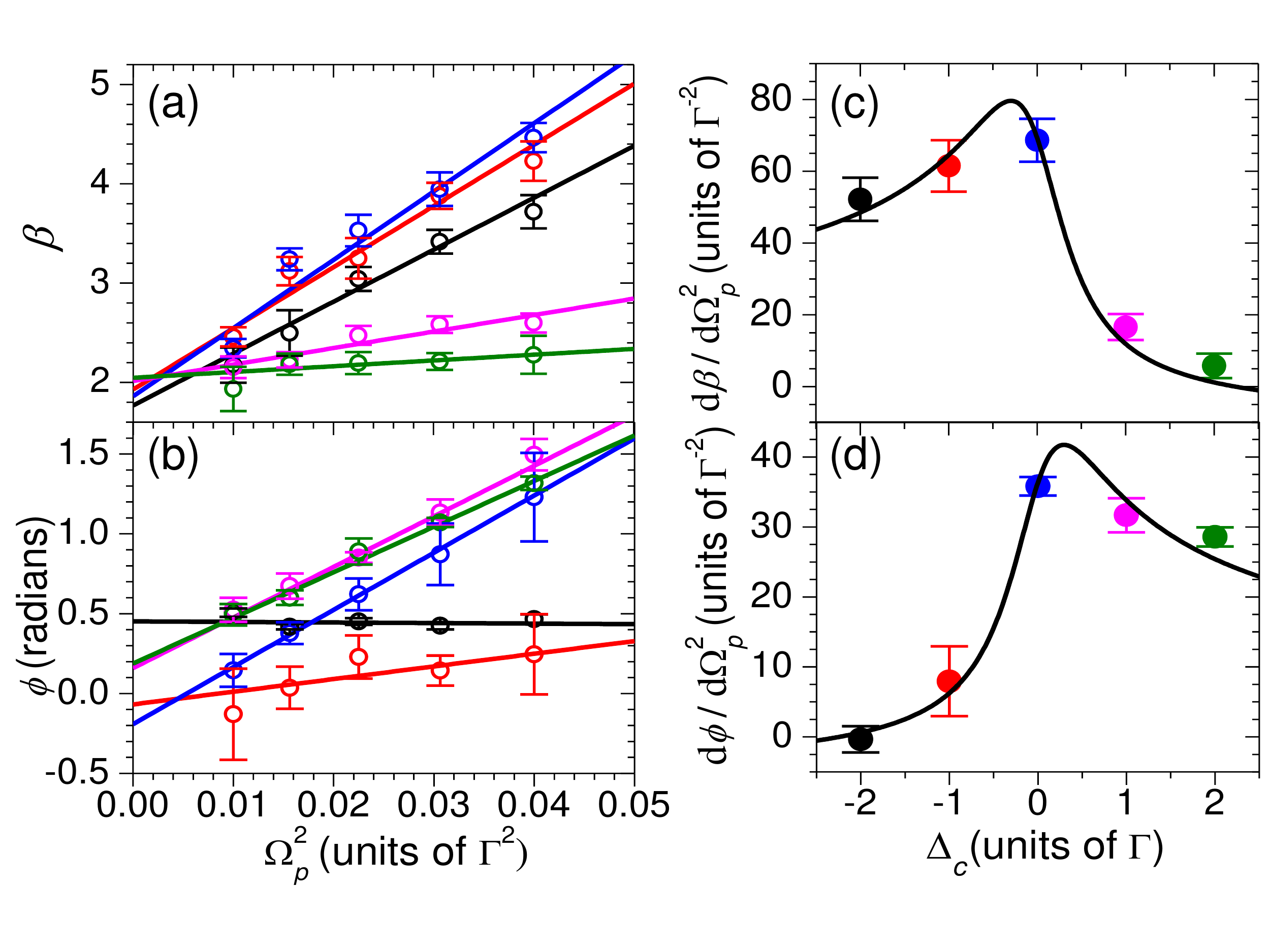}}
	\caption{Attenuation coefficient and phase shift due to the dipole-dipole interaction (DDI). (a,b) Attenuation coefficient $\beta$ and phase shift $\phi$ at $\delta = 0$ as functions of $\Omega_p^2$ in the presence of the DDI. Circles are the experimental data taken with $\alpha$ (optical depth) = 81$\pm$3, $\Omega_c = 1.0$$\Gamma$, and $\Delta_c = -2$$\Gamma$ (black), $-1$$\Gamma$ (red), 0 (blue), 1$\Gamma$ (magenta), and 2$\Gamma$ (olive), where $\Gamma$ = 2$\pi$$\times$6 MHz. Straight lines are the best fits. (c,d) Circles are data points of the slope of straight line in (a,b) versus the corresponding $\Delta_c$. Black lines are the best fits, which  determine $S_{\rm DDI}$ = 37/$\Gamma$$^{3/2}$ in (c) and 38/$\Gamma$$^{3/2}$ in (d). The error bars represent the standard deviation of measured values.
}
	\label{fig:DDI}
	\end{figure*}
}
\newcommand{\FigThree}{
	\begin{figure}[t]
	\center{\includegraphics[width=87mm]{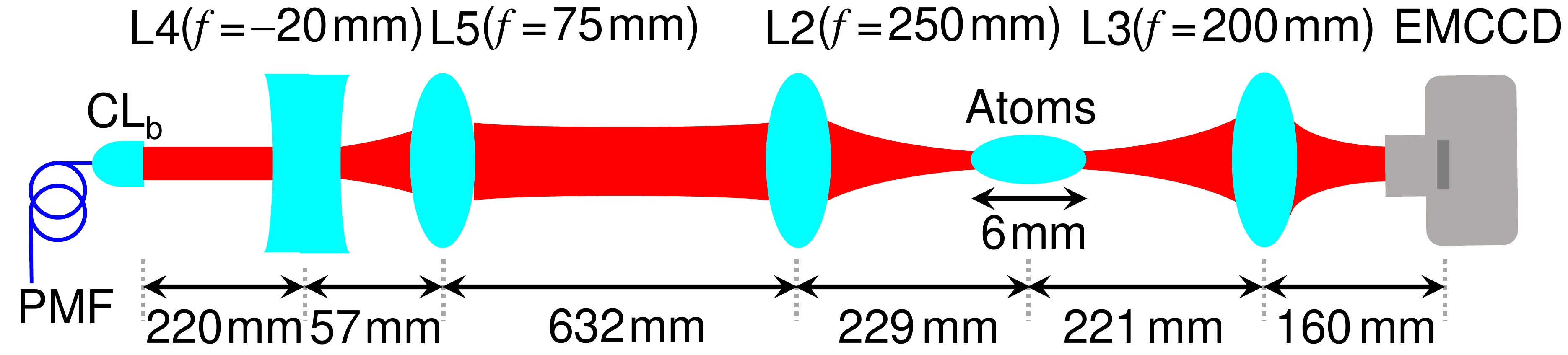}}
	\caption{Sketch of key elements for taking images of the probe beam. Values of $f$ indicate the focal lengths of lenses. PMF: polarization-maintained optical fiber; CL$_{b}$: collimation lenses; L2 and L3: lenses; EMCCD: electron-multiplying charge-coupled device camera (Andor DL-604M-OEM). The probe beam coming out of CL$_{b}$ was collimated and had the e$^{-1}$ full width of 0.92~mm. We adjusted the separation between L4 and L5 such that the beam was focused on nearly the center of the atomic cloud by L2, and had the width of 39~$\mu$m at the focal point.
}
	\label{fig:ImageSystem}
	\end{figure}
}
\newcommand{\FigFour}{
	\begin{figure}[t]
	\center{\includegraphics[width=55mm]{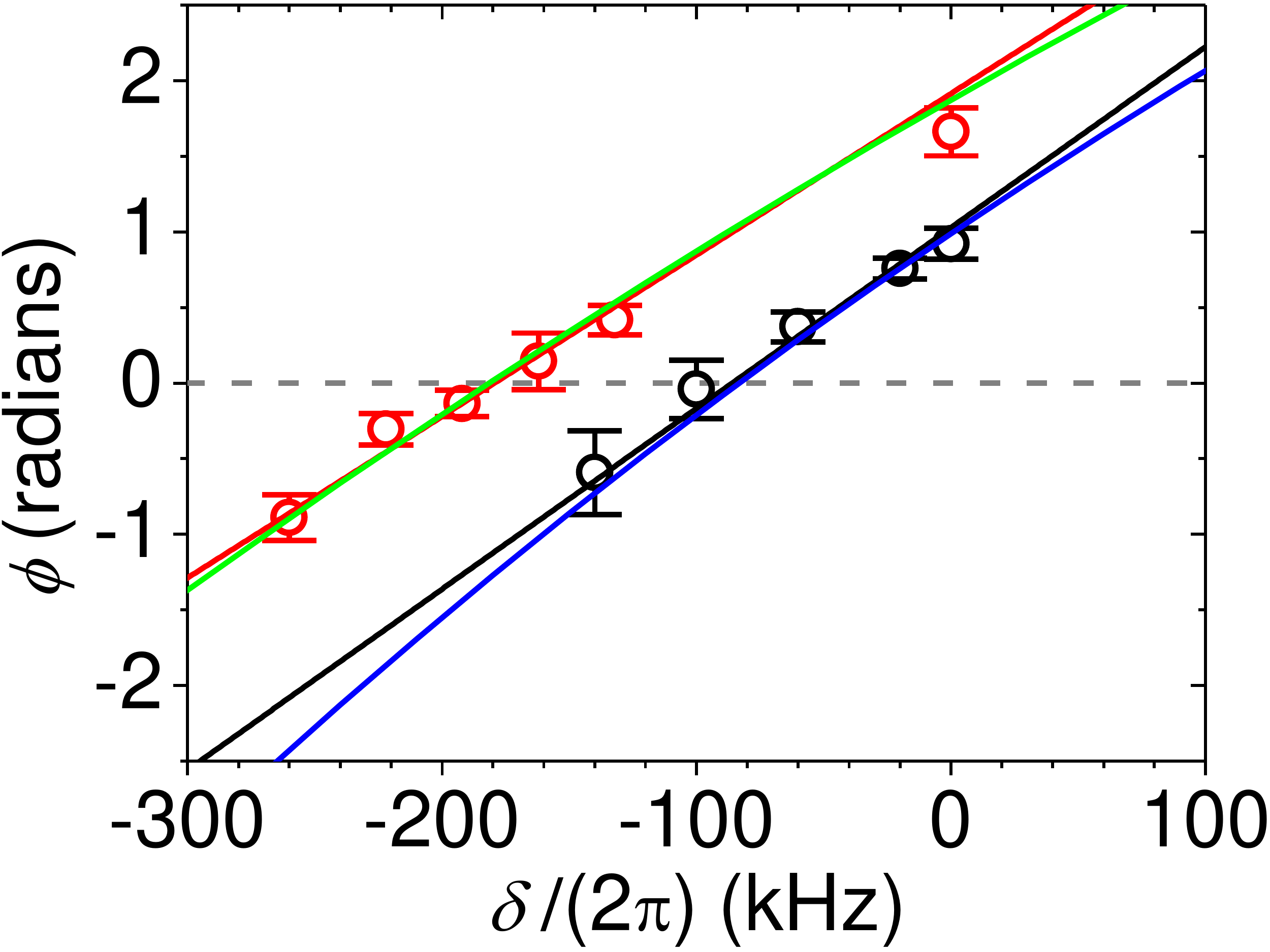}}
	\caption{Determination of the two-photon detuning for the zero phase shift. Black (or red) circles are the experimental data of phase shift versus two-photon detuning measured at $\alpha$ (optical depth) = 82, $\Omega_c =$ 1.0$\Gamma$, $\Delta_c =$ 1.0$\Gamma$, and $\Omega_p =$ 0.10$\Gamma$ (or 0.20$\Gamma$). Black and red lines represent the linear best fits. We determined the two-photon detuning, $\delta_0$, for the zero phase shift, i.e. $\phi=0$, by the intersection between the best fit and the gray dashed line. As references, blue and green lines are the theoretical predictions calculated numerically\ncite{OurMFT}. The error bars represent the standard deviation of measured values.
}
	\label{fig:phi_versus_delta}
	\end{figure}
}
\newcommand{\FigFive}{
	\begin{figure}[t]
	\center{\includegraphics[width=86mm]{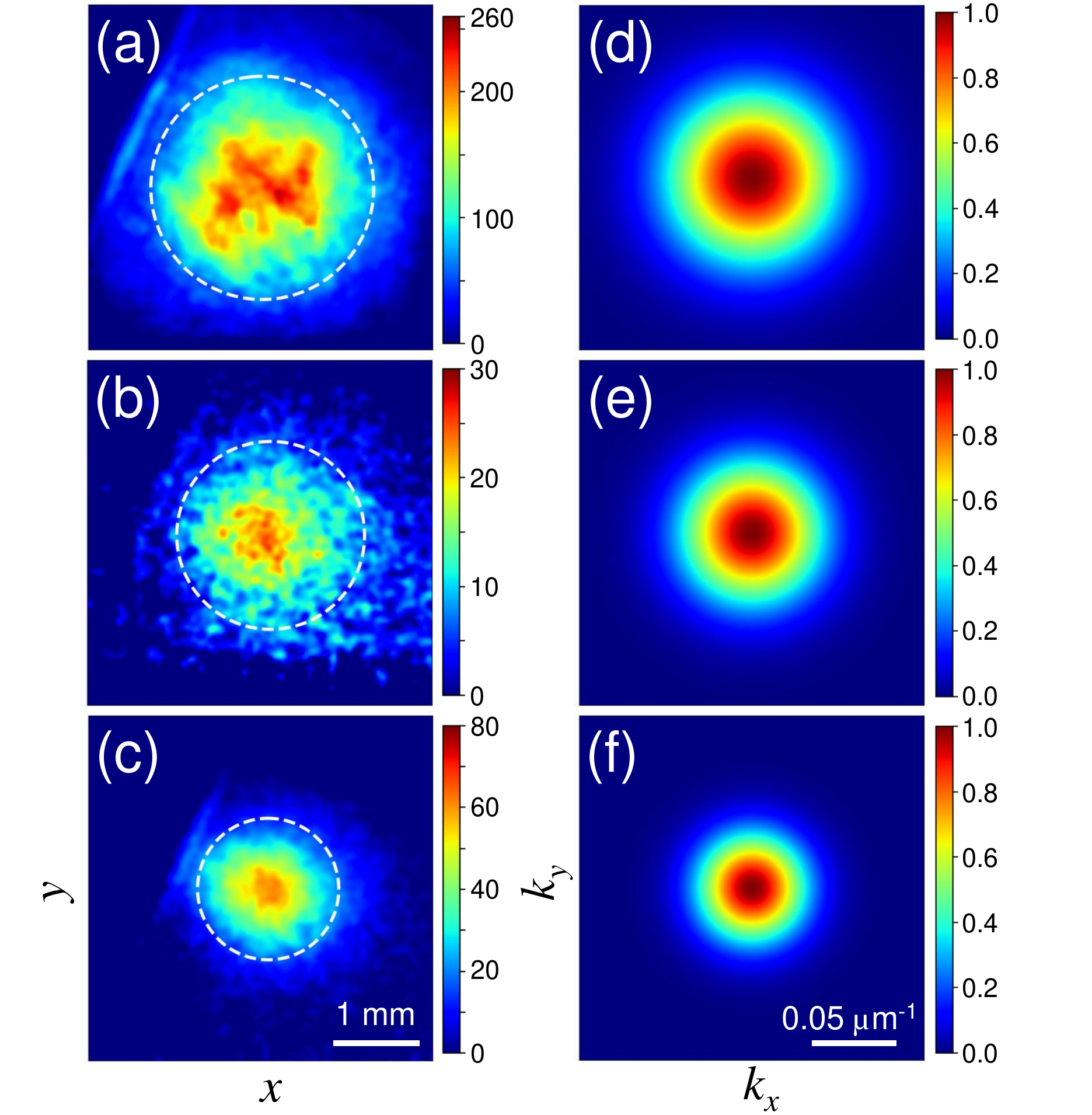}}
	\caption{Images of the output probe beam profile and transverse momentum distributions of the Rydberg polaritons. (a-c) Images of the probe beam profile taken by the electron-multiplying charge-coupled device camera (EMCCD) at $\Omega_p = 0.2$$\Gamma$ in the absence of the atoms, and at $\Omega_p = 0.1$$\Gamma$ and $0.2$$\Gamma$ in the presence of the atoms, respectively. Color represents the gray level detected by the EMCCD. We fit each image with a Gaussian function, and draw a white-dashed-line circle of the diameter equal to the e$^{-1}$ full width of the best fit. The diameters of the circles from top to bottom are 2.6, 2.2 and 1.6~mm. (d-f) Transverse momentum distributions of the probe photons at the output, i.e., the Rydberg polaritons in the atomic cloud, derived from the best fits of the images in (a-c), respectively. Color represents the normalized probability density. The e$^{-1}$ full widths of the distributions from top to bottom are 0.10, 0.087 and 0.065~$\mu$m$^{-1}$.
}
	\label{fig:Thermalization}
	\end{figure}
}
\newcommand{\FigSix}{ 
	\begin{figure}[t]
	\center{\includegraphics[width=60mm]{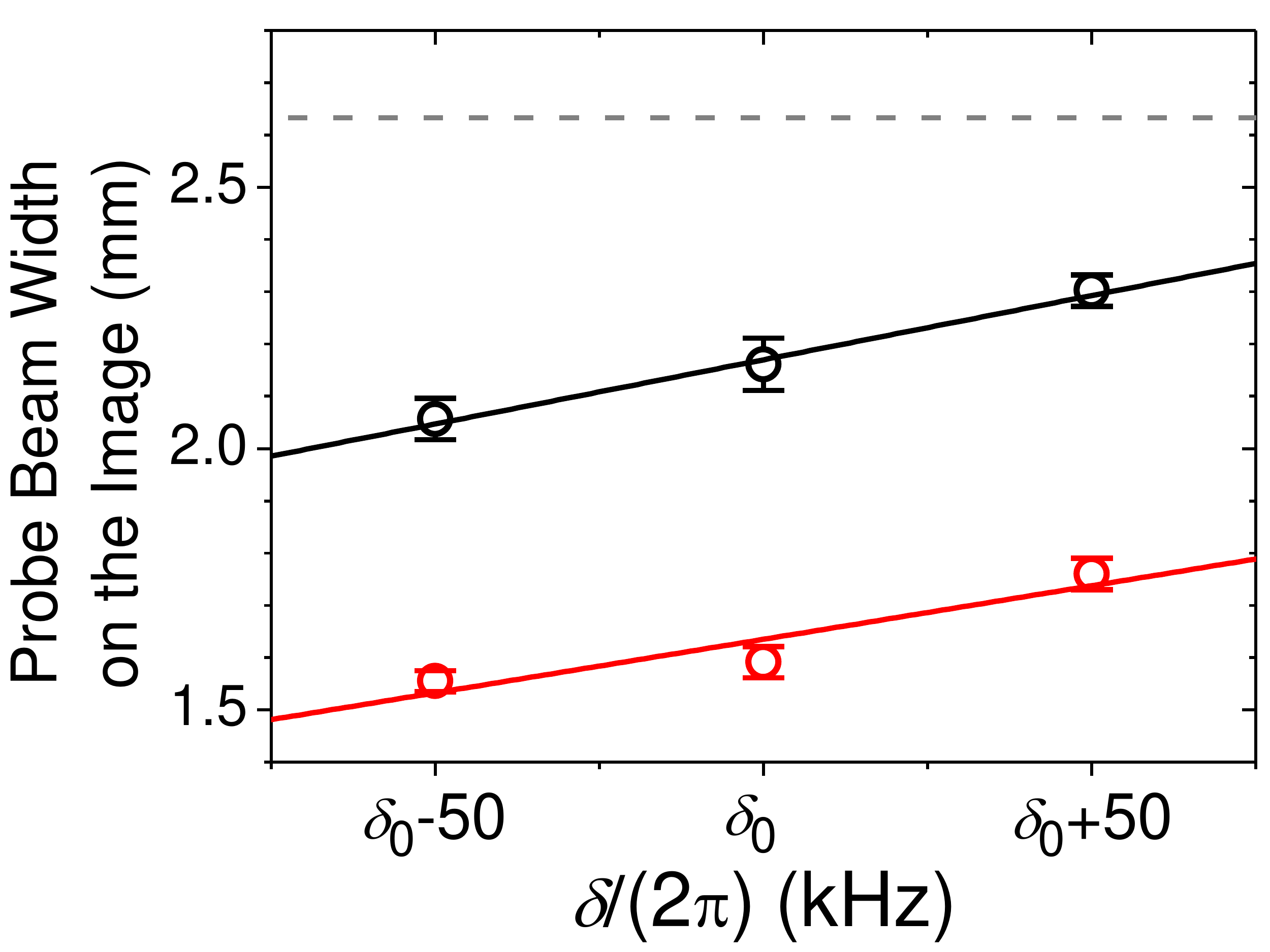}}
	\caption{The output probe beam width as a function of the two-photon detuning. Black and red circles are the experimental data of the e$^{-1}$ full width of the output probe beam versus the two-photon detuning. The black (or red) data points were taken at $\alpha$ (optical depth) = 82, $\Omega_c =$ 1.0$\Gamma$, $\Delta_c = 1.0\Gamma$, and $\Omega_p =$ 0.10$\Gamma$ (or 0.20$\Gamma$). Black and red lines are the linear best fits of the experimental data, and their slopes are 2.5~$\mu$m/kHz and 2.1~$\mu$m/kHz, respectively. Gray dashed line indicates the probe beam width measured without the presence of the atoms, i.e., the input probe beam width. The error bars represent the standard deviation of measured values.
}
	\label{fig:beam_size}
	\end{figure}
}
\newcommand{\FigSeven}{ 
	\begin{figure}[t]
	\center{\includegraphics[width=87mm]{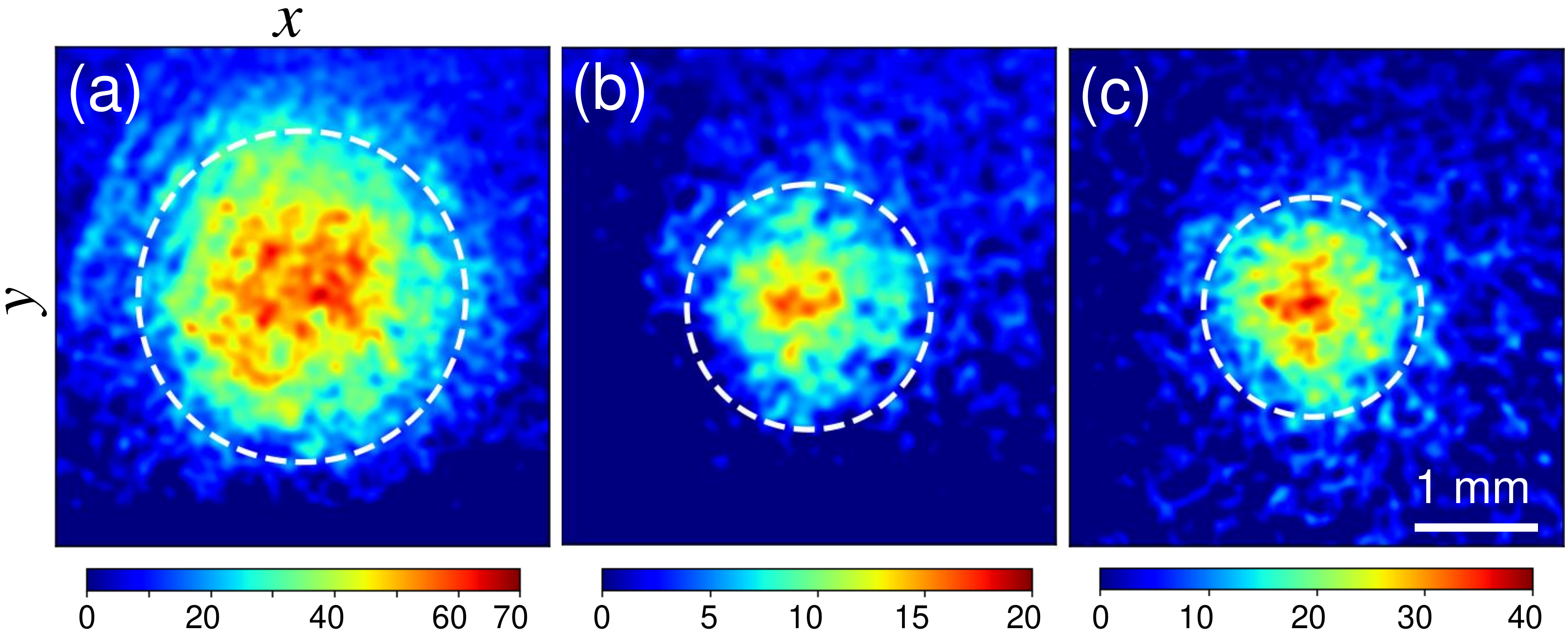}}
	\caption{Images of the output probe beam profile taken at $\Delta_c =$ $-1$$\Gamma$. Here, the phase shift induced by the dipole-dipole interaction is minimal. (a) The image was taken in the absence of the atoms and at $\Omega_p =$ 0.1$\Gamma$. (b, c) The images were taken in the presence of the atoms, and $\Omega_p =$ 0.1$\Gamma$ and 0.2$\Gamma$, respectively. Color represents the gray level detected by the electron-multiplying charge-coupled device camera. We fit each image with a Gaussian function, and draw a white-dashed-line circle of the diameter equal to the e$^{-1}$ full width of the best fit. The diameters of the circles in (a), (b) and (c) are 2.6, 2.0 and 1.8~mm, respectively.
}
	\label{fig:Thermalization_negative}
	\end{figure}
}

\section{Introduction}

\indent Rydberg-state atoms provide a strong dipole-dipole interaction (DDI) and a dipole blockade mechanism\ncite{blockade_Zoller2000, blockade_Gould2004, blockade_Pfau2007, SaffmanRMP} suitable for applications in quantum information processing, such as quantum simulators\ncite{simulator_Zoller2010, simulator_Lukin2018, simulator_Ahn2018}, quantum logic gates\ncite{gate_Browaeys2010, gate_Saffman2010, gate_Lukin2019, gate_Rempe2019}, and single-photon sources\ncite{SP_Saffman2002, SP_Kuzmich2016, SP_Pfau2018}. The effect of electromagnetically induced transparency (EIT) reduces the propagation speed of light in media and significantly increases the light-matter interaction time\ncite{EIT_Fleischhauer2005, EIT_OurPRL2006, SLP_OurPRL2012, QMemory_OurPRL2013, EIT_YCChen2018}. Thus, the EIT effect involving the Rydberg atoms can efficiently mediate significant photon-photon interactions\ncite{REIT_Adams2010, REIT_Fleischhauer2011, REIT_Lukin2012, REIT_Fleischhauer2015, REIT_Hofferberth2016, XPM_Rempe2016}, and this issue has attracted a great deal of attention recently. Most of the previous Rydberg-EIT studies concentrated on strongly-correlated photons in the dipole blockade regime, which emerges for the strongly-interacting Rydberg atoms with a high principal quantum number, $n$, close to 100. This leads to, for example, the two-body photon-photon gates\ncite{XPM_Rempe2016, gate_Rempe2019} and strongly-correlated many-body phases\ncite{REIT_Lukin2012, REIT_Fleischhauer2015}.

The system of the Rydberg EIT polaritons is an ensemble of bosonic quasiparticles\ncite{DSP_Juzeliunas2002, DSP_Fleischhauer2002}. The dispersion relation and momentum distribution of the Rydberg polaritons can be associated with their effective mass and the temperature\ncite{DSP_BEC_Fleischhauer2008}. The phase shift and attenuation of the output light induced by the DDI can be viewed as a consequence of the elastic and inelastic collisions among the polaritons. Hence, the change rates of the phase shift and attenuation are related to the elastic and inelastic collision rates, respectively. Finally, the propagation time of slow light or the diffusion time of stationary light in the EIT system is exactly the interaction time of the particles.

\FigOne

Here we proposed and experimentally demonstrated an innovative idea of utilizing a medium with a high optical depth (OD) and Rydberg atoms with a low principal quantum number to create a weakly-interacting many-body system based on the EIT effect. In this work, OD $\approx 80$ and $n = 32$ were used. We studied the EIT polaritons also known as the dark-state polaritons\ncite{DSP_Fleischhauer2000, DSP_Juzeliunas2002, DSP_Fleischhauer2002}, which, in the case of the Rydberg EIT, are quasi-particles representing superpositions of photons and Rydberg coherences. The Rydberg coherence is the coherence between the atomic ground and Rydberg states. Weakly-interacting Rydberg polaritons under an OD-enhanced interaction time can be employed in the study of many-body physics, such as the Bose-Einstein condensation (BEC)\ncite{BEC_Cornell1995, MagnonBEC_nature2006, EPBEC_nature2006, EPBEC_science2007, BEC_Ketterle1995, EPBEC_RMP2010, DSP_BEC_Fleischhauer2008, PhotonBEC_nature2010}.

The low-$n$ Rydberg state, together with a moderate polariton density, ensured the weak-interaction regime in our experiment, i.e. $r_B^3/r_a^3\ll 1$, where $r_B$ is the blockade radius and $r_a$ is the half mean distance between Rydberg polaritons. In this regime, the mean number of Rydberg polaritons in a blockade sphere is much less than the unity, and thus attenuation due to the blockade effect occurs rarely. The collision cross section of polaritons due to the DDI is around $\mu$m$^2$ in the Rydberg state of $n = 32$, leading to the elastic collision rate of a few MHz at the Rydberg-polariton density of 2$\times$$10^9$ cm$^{-3}$ used in this work. On the other hand, the high-OD medium provides for the Rydberg polaritons a long interaction time of a few $\mu$s, as demonstrated here and in our previous works\ncite{SLP_OurPRL2012, QMemory_OurPRL2013}. Therefore, a high OD can result in a sufficiently large product of the collision rate and interaction time for weakly-interacting Rydberg polaritons, enabling them to exhibit many-body phenomena\ncite{DSP_BEC_Fleischhauer2008, BEC_Ketterle1998, RDSP_Fleischhauer2013}. 

In this work, we systematically studied the phase shift and attenuation of an output probe light for the EIT involving a Rydberg state of $|32D_{5/2}\rangle$. Owing to the high OD and low intrinsic decoherence rate in the system, to the best of our knowledge, our experiment is the first one to demonstrate the DDI effect in the truly weak-interaction regime, i.e., $r_B^3/r_a^3 <0.1$, as compared with previous experiments\ncite{att1, REIT_Adams2010, REIT_Lukin2012, att2, att3, spectralshift, XPM_Rempe2016, phase1, phase2}. We were able to clearly observe the DDI-induced attenuation and phase shift even at $r_B^3/r_a^3= 0.02$. Furthermore, we varied the DDI strength via the input photon flux and measured the transverse momentum distribution of the Rydberg polaritons. A larger DDI strength caused the width of the momentum distribution to become notably smaller, indicating the thermalization process was driven by elastic collisions\ncite{DSP_BEC_Fleischhauer2008, Thermalization2010, Thermalization2013}. The observed reduction of the momentum distribution width suggested that the combination of the $\mu$m-range DDI strength and the $\mu$s-long light-matter interaction time can make the BEC of the Rydberg polaritons feasible.

\vspace*{\baselineskip}
\section{Results}

\vspace*{0.5\baselineskip}
\subsection{Experimental setup.} We carried out the experiment in cold $^{87}$Rb atoms produced by a magneto-optical trap (MOT). There were typically 5$\times$$10^8$ trapped atoms with a temperature of about 350~$\mu$K in the MOT\ncite{REIT_OurPRA2019, OurCigarMOT}. The cigar-shaped atomic cloud had the dimension of 1.8$\times$1.8$\times$6.0~mm$^{3}$. Before each measurement, we first turned off the magnetic field of the MOT and then performed the dark MOT for 2.5~ms to increase the optical depth (OD) of the system\ncite{QMemory_OurPRL2013, OurSSL}. The OD of the atomic cloud was 81$\pm$3. When the experimental condition of a low OD was needed, we were able to adjust the dark-MOT parameter to reduce the OD of the system by about 4 folds.

Figure~\ref{fig:EnergyLevel}(a) shows the transitions driven by the probe and the coupling fields in the EIT system. We optically pumped all population to a single Zeeman state $|F=2,m_F=2\rangle$ at the ground level $|5S_{1/2}\rangle$\ncite{OurSSL}. Since the probe and coupling fields were $\sigma_+$-polarized, each of the energy levels $|1\rangle$, $|2\rangle$ and $|3\rangle$ shown in Fig.~\ref{fig:EnergyLevel}(a) was a single Zeeman state. The spontaneous decay rate, $\Gamma$, of $|3\rangle$ is 2$\pi$$\times$6~MHz, and $|1\rangle \leftrightarrow |3\rangle$ is a cycling transition. The life time, $\Gamma_2^{-1}$, of $|2\rangle$ is about 30~$\mu$s\ncite{RydbergStateLifeTime}, and $\Gamma_2 \ll \gamma_0$ in this work, where $\gamma_0$ is the intrinsic decoherence rate of the system. Thus, the influence of $\Gamma_2$ was negligible. We estimated $n_{\rm atom} \approx$ 5$\times$10$^{10}$~cm$^{-3}$ in the experiment, and $C_6 = $ $-2\pi$$\times$260~MHz$\cdot\mu$m$^6$ for $|32D_{5/2}, m_J = 5/2\rangle$ by using $D_{\varphi}$ = 0.343, where $D_{\varphi}$ is the coefficient resulted from the angular momentum states of two interacting Rydberg atoms being $|J=5/2,m_J = 5/2\rangle$\ncite{Saffman2008}. Such $C_6$ together with $n_{\rm atom}$ can make the DDI effect unobservable under a small $\Omega_p$ or a low OD\ncite{REIT_Adams2010, dipolardephasing, adams2008, spectralshift}. 

\FigTwo

Figure~\ref{fig:ExpSetup}(b) shows the experimental setup. The probe field came from the first-order beam of an acousto-optic modulator (AOM) and was coupled to a polarization maintained fiber (PMF). The PMF delivered the probe field to the cold atoms. The probe field was twice diffracted from the AOM in the double-pass scheme. The AOM was used to shape the probe pulse, and its driving rf frequency and amplitude were precisely tuned or varied by a function generator. We used a rubidium atomic standard (SRS FS752) as the external clock of the function generator. The blue path in Fig.~1(b) represents the coupling field, which was switched on and off with another AOM not drawn in the figure. Inside the atom cloud, the probe and coupling fields counter-propagated and completely overlapped to minimize the Doppler effect. The e$^{-1}$ full widths of the probe and coupling beams were  130 and 250~$\mu$m, respectively. After passing through the atoms, the probe field was detected by a photomultiplier tube (PMT). A digital oscilloscope (Agilent MSO6014A) acquired the signal from the PMT and produced raw data.

The entire setup of red and brown optical paths shown in Fig.~\ref{fig:ExpSetup}(b) formed the beat-note interferometer, which measured the phase shift of the output probe field (see beat-note interferometer in method section). The brown optical paths were present only in the phase shift measurement. 

\vspace*{0.5\baselineskip}
\subsection{Theoretical model.}
 The mean field model developed in Ref.~\rcite{54}{OurMFT}, which describes the DDI-induced attenuation and phase shift in the system of weakly-interacting Rydberg polaritons, is summarized in this section. Rydberg polaritons are randomly distributed and can be considered approximately as particles of a nearly ideal gas due to their weak interaction. Thus, the nearest-neighbor distribution\ncite{NND} was utilized in our theory. Using the probability function of the nearest-neighbor distribution and the atom-light coupling equations of an EIT system, we derived the following analytical formulas of the steady-state DDI-induced attenuation coefficient $\Delta\beta$ and phase shift $\Delta\phi$:
\begin{eqnarray}
	\Delta\beta &=&
		2 S_{\rm DDI} \left( \frac{\sqrt{W_c-2\Delta_c}}{W_c} 
		-\frac{3\gamma_0\sqrt{W_c+2\Delta_c}}{\Omega_c^2} \right)
		\Omega_p^2,~~
\label{eq:Delta_beta} \\ 
	\Delta\phi &=&
		S_{\rm DDI} \left( \frac{\sqrt{W_c+2\Delta_c}}{W_c} 
		-\frac{3\gamma_0\sqrt{W_c-2\Delta_c}}{\Omega_c^2}  \right)
		\Omega_p^2,~
\label{eq:Delta_phi}
\end{eqnarray}
\vspace*{-\baselineskip}
\begin{eqnarray}
	S_{\rm DDI} &\equiv& 
		\frac{\pi^2\alpha\Gamma\sqrt{|C_6|} n_{\rm atom} \varepsilon}{3\Omega_c^3},
\label{eq:S_DDI} \\
	W_c &\equiv& \sqrt{\Gamma^2+4\Delta_c^2},
\label{eq:W_c}
\end{eqnarray}
where $\alpha$ is the OD of the system, $\Gamma$ is the spontaneous decay rate of the excited state $|3\rangle$, $\gamma_0$ is the intrinsic decoherence rate in the system, $C_6$ is the van der Waals coefficient in SI units of Hz$\cdot$m$^6$, $n_{\rm atom}$ is the atomic density, and $\varepsilon$ is a phenomenological parameter. In the above equations, we assume $\Omega_p^2 \ll \Omega_c^2$ and $r_B^3 \ll r_a^3$, and consider the case that the two-photon detuning $\delta$ is much smaller than the EIT linewidth and its influence on $\Delta\beta$ and $\Delta\phi$ is negligible. Furthermore, $\varepsilon$ is utilized to relate the average Rydberg-state population ($\rho_{22})$ to the input probe and coupling Rabi frequencies ($\Omega_p$ and $\Omega_c$) as $\rho_{22}= \varepsilon\Omega_p^2/\Omega_c^2$. Equations~(\ref{eq:Delta_beta}) and (\ref{eq:Delta_phi}) are for the case of $C_6 <0 $. They qualitatively agree with those in the pioneering work of Ref.~\rcite{56}{Phol2011}, and served as the theoretical references for the experimental data in this work. 

The attenuation coefficient $\beta$ and phase shift $\phi$ of the probe field are defined by the output-input ($\Omega'_p$-$\Omega_p$) relation of $\Omega'_p = \Omega_p \exp(i\phi -\beta/2)$. Thus, 
\begin{eqnarray} 
	\beta &=& \beta_0 + \Delta\beta 
		\approx \frac{2\alpha\gamma_0\Gamma}{\Omega_c^2} + \Delta\beta, 
\label{eq:beta} \\
	\phi &=& \phi_0 + \Delta\phi
		\approx \frac{\alpha\Gamma\delta}{\Omega_c^2} + \Delta\phi, 
\label{eq:phi}
\end{eqnarray}
where $\beta_0$ and $\phi_0$ are the attenuation coefficient and phase shift in the absence of DDI\ncite{OurMFT}, $\delta (= \Delta_p + \Delta_c)$ is the two-photon detuning. The mean field theory predicted that both $\beta$ and $\phi$ depend on $\Omega_p^2$ linearly, and the dependence of $\Omega_p^2$ comes from the Rydberg-state population $\rho_{22}$. In addition, the slope of $\beta$ or $\phi$ versus $\Omega_p^2$ is asymmetric with respect to $\Delta_c = 0$.

\vspace*{0.5\baselineskip}
\subsection{DDI-induced attenuation and phase shift.}
 We first verified that the DDI effect can be observed in the system consisting of weakly-interacting Rydberg polaritons. The attenuation coefficient $\beta$ and phase shift $\phi$ were measured as functions of the square of the probe Rabi frequency $\Omega_p^2$, as shown in Figs.~\ref{fig:DDI}(a) and \ref{fig:DDI}(b). A nonzero two-photon detuning, $\delta$, can significantly affect the phase shift and attenuation of the output probe light in the EIT medium. Thus, we utilized a beat-note interferometer to carefully determine the probe frequency for $\delta = 0$ (see beat-note interferometer in method section). The uncertainty of $\delta/2\pi$ was $\pm$30 kHz, including the accuracy of the beat-note interferometer of $\pm$10 kHz and the long-term drift of the two-photon frequency of $\pm$20~kHz. 

At $\delta = 0$, we applied a square input probe pulse and measured the steady-state attenuation coefficient $\beta$ and phase shift $\phi$ of the output probe pulse. The measurement procedure and representative data can be found in Supplementary Note 3. Figure~\ref{fig:DDI}(a) [and \ref{fig:DDI}(b)] shows $\beta$ (and $\phi$) versus $\Omega_p^2$ at various $\Delta_c$, where $\Omega_p^2$ corresponds to the peak or center intensity of the input Gaussian beam. At a given $\Delta_c$, the data points approximately formed a straight line, as expected from the theory. We fitted the data with a linear function and plotted the slope of the best fit against $\Delta_c$, as shown in Fig.~\ref{fig:DDI}(c) [and \ref{fig:DDI}(d)]. It can be noticed that the $y$-axis interceptions of best fits were scattered in Fig.~\ref{fig:DDI}(b). Since a change of 30~kHz in $\delta/2\pi$ resulted in that of 0.4 rad in phase, the uncertainty of $\delta$ made the interceptions scatter around zero.

The strengths of the DDI effect on $\beta$ and $\phi$, i.e., the slope of $\beta$ versus $\Omega_p^2$ and that of $\phi$ versus $\Omega_p^2$ (denoted as $\chi_{\beta}$ and $\chi_{\phi}$), were asymmetric with respect to $\Delta_c = 0$ as shown in Figs.~\ref{fig:DDI}(c) and \ref{fig:DDI}(d). Such asymmetries are expected from the theory, as demonstrated by Eqs.~(\ref{eq:Delta_beta}) and (\ref{eq:Delta_phi}). Considering the coupling detunings of $-|\Delta_c|$ and $|\Delta_c|$, $\chi_{\beta}$ (or $\chi_{\phi}$) of $-|\Delta_c|$ was always larger (or smaller) than that of $|\Delta_c|$. The physical picture of the asymmetries is discussed in Ref.~\rcite{54}{OurMFT}. We fitted the data points in Figs.~\ref{fig:DDI}(c) and \ref{fig:DDI}(d) with the functions of $\chi_{\beta}$ and $\chi_{\phi}$, where $\gamma_0$ was set to 0.012$\Gamma$, and $S_{\rm DDI}$ was the only fitting parameter. The $S_{\rm DDI}$ of the best fit shown in Fig.~\ref{fig:DDI}(c) is in good agreement with that of Fig.~\ref{fig:DDI}(d). Using Eq.~(\ref{eq:S_DDI}) and the values of $S_{\rm DDI}$, $C_6$, $n_{\rm atom}$ and $\Omega_c$, we obtained $\varepsilon$ = 0.43$\pm$0.05. Considering that $\varepsilon$ related the average Rydberg-state population to $\Omega_p^2$ and that $\Omega_p^2$ corresponded to the center intensity of input Gaussian beam, the measured value of $S_{\rm DDI}$ is also reasonable. Thus, the experimental data are in the good agreement with the theoretical predictions, revealing that the low-$n$ Rydberg polaritons in the high-OD medium can form a weakly-interacting many-body system.

\vspace*{0.5\baselineskip}
\subsection{Feasibility of thermalization process.}
 To observe the thermalization process in this weakly-interacting system, we varied the DDI strength and measured the transverse momentum distribution of the Rydberg polaritons. The moveable mirror M$_{b}$ shown in Fig.~\ref{fig:ExpSetup}(b) was installed to direct the output probe beam to the EMCCD. Figure~\ref{fig:ImageSystem} shows the key elements for the EMCCD image of the probe beam. We reduced the waist of the input probe beam by adding the lens pair of L4 and L5, as shown in the figure, and made the input photons or initial Rydberg polaritons have a large transverse momentum distribution. In the measurements of Figs.~\ref{fig:phi_versus_delta}-\ref{fig:Thermalization_negative}, the e$^{-1}$ full width of the probe beam at the center of the atomic cloud was reduced to 39~$\mu$m, while we kept the coupling beam size the same as before. Since the coupling beam size was 6.4 times larger than the probe beam size, the waveguide effect due to the coupling intensity profile should play little role in the data of these figures. Furthermore, the attenuation of Rydberg polaritons decreases the particle density and consequently the collision rate. While the propagation delay time, i.e., the interaction time, at $\Delta_c$ of a small magnitude can be maintained approximately the same, the DDI-induced attenuation coefficient of $\Delta_c =$ +1$\Gamma$ is significantly smaller than those of $\Delta_c =$ 0 and $-1$$\Gamma$. Hence, we chose the experimental condition of $\Delta_c =$ +1$\Gamma$ in the study.

\FigThree

Following the idea in Ref.~\rcite{28}{DSP_BEC_Fleischhauer2008}, we estimate the elastic collision rate, which is responsible for the thermalization, of the Rydberg polaritons from the formula below:
\begin{equation}
\label{eq:collision_rate}
	\frac{d\phi}{dt} = R_c \phi_c,
\end{equation}
where $d\phi/dt$ is the phase shift per unit time, $R_c$ denotes the collision rate, and $\phi_c$ represents the phase shift per collision. Note that in the case of $\Delta_c =$ $-1$$\Gamma$, for example, some collisions result in positive phase shifts and some result in negative phase shifts\ncite{OurMFT}, and thus the net phase shift is close to zero. Such a case cannot be used to make an estimation of the collision rate. To ensure that the phase shifts induced by the collisions are all positive, we should use the case of $\Delta_c =$ 0 or $+1$$\Gamma$ in the estimation. The value of $d\phi/dt$ can be estimated from the experimentally observed value of $\Delta\phi/\tau_{d}$, where $\Delta\phi$ is the DDI-induced phase shift measured at the output of the medium and $\tau_{d}$ is the propagation delay time, i.e., the interaction time of Rydberg polaritons. Using $\tau_d =$ 2.1~$\mu$s and the data point shown in Fig.~\ref{fig:DDI}(b) of $\Delta_c = +1$$\Gamma$ and $\Omega_p = 0.2$$\Gamma$ resulted in $d\phi/dt =$ 0.64~rad/$\mu$s. Note that the experimental condition of $\Omega_p = 0.2$$\Gamma$ corresponded to the Rydberg-polariton density of 2$\times$10$^9$~cm$^{-3}$. Considering the hard-sphere collision, $\phi_c = \bar{k} a$ where $\hbar \bar{k}$ is the root-mean-square value of the relative momentum between two colliding bodies, and $a$ is the radius of a hard sphere. In view of the Rydberg polaritons, $a$ can be treated as the blockade radius\ncite{HardSphere}, which was about 2.1~$\mu$m in our case according to the formula\ncite{REIT_Lukin2012} of $r_B=(2C_{6} \Gamma /\Omega_c^2)^{1/6}$. The momentum distribution, which we measured in this study and will be shown later, indicated $\bar{k} = 0.051$ $\mu$m$^{-1}$. The values of $\bar{k}$ and $a$ result in $\phi_c$ = 0.11~rad. Thus, we obtain $R_c =$ 6.0~MHz by inserting the values of $d\phi/dt$ and $\phi_c$ in Eq.~(\ref{eq:collision_rate}). Under such an elastic collision rate, it was feasible to observe the thermalization effect in our experiment.

\vspace*{0.5\baselineskip}
\subsection{Procedure of the beam profile measurement.}
 We measured the output probe beam size on the EMCCD at $\Delta_c =$ 1$\Gamma$. To avoid the lensing effect\ncite{eif}, we set the two photon detuning, $\delta_0$, corresponding to the zero phase shift, giving $\phi = 0$ in the measurement. That is, at the two-photon resonance ($\delta = 0$) there is a positive DDI-induced phase shift ($\phi > 0$), and we deliberately set a negative two-photon detuning, $\delta_0$, to eliminate the phase shift, i.e., $\phi = 0$ at $\delta = \delta_0$ where $\delta_0/2\pi$ was equal to $-85$ or $-180$ kHz in the measurement of $\Omega_p =$ 0.1$\Gamma$ or 0.2$\Gamma$, respectively. Please note $\phi = 0$ is the condition that the phase shift due to the ordinary EIT effect at $\delta_0$ cancels out the phase shift due to the DDI effect. Details of the measurement of the probe beam size, i.e., the transverse momentum distribution of the Rydberg polaritons, and those of the study on the lensing effect can be found in Supplementary Notes 4 and 5.

\FigFour
 
The procedure of taking the image of the output probe beam profile is summarized in the following steps: (i) The experimental parameters were determined and set to the designed values. (ii) The probe frequency for $\delta = 0$ was determined by the method of the beat-note interferometer. (iii) At the above experimental parameters, we measured $S_{\rm DDI}$, and confirmed that the measured value was consistent with the $S_{\rm DDI}$ shown in Fig.~\ref{fig:DDI}. (iv) The two-photon detuning, $\delta_0$, corresponding to $\phi=0$ was determined. Representation data are shown in Fig.~\ref{fig:phi_versus_delta}. We fitted the data with a straight line, and the best fit determined $\delta_0$. The uncertainty of $\delta_0$ was $\pm$2$\pi\times$30~kHz. (v) At $\delta_0$, we took images of the output probe beam profile. We repeated the steps (iv) and (v) for different values of $\Omega_p$. (vi) The value of $S_{\rm DDI}$ was measured again to verify it was unchanged during the steps (iv) and (v). 

\FigFive

\vspace*{0.5\baselineskip}
\subsection{Images of the probe beam profile.}
Figure~\ref{fig:Thermalization}(a) shows the probe beam profile in the absence of the atom, while Figs.~\ref{fig:Thermalization}(b) and \ref{fig:Thermalization}(c) show those in the presence of the atoms. It can be clearly observed that a larger value of $\Omega_p^2$, i.e., a higher density of Rydberg polaritons because of $n_{\rm atom}\rho_{22} \propto \Omega_p^2$, caused the beam size to become smaller. Since transverse momentums of the Rydberg polaritons were carried by the probe photons leaving the medium, the intensity profile of the output probe beam was able to be used to derive the transverse momentum distribution\cite{EPBEC_nature2006, DSP_BEC_Fleischhauer2008}. We fitted the intensity profiles of three images with a two-dimension Gaussian function, and utilized the results of the best fits to construct the momentum distributions shown in Figs.~\ref{fig:Thermalization}(d)-\ref{fig:Thermalization}(f). 
The relation between the transverse momentum distribution of photons at the atom cloud and the intensity profile of an EMCCD image can be found in Supplementary Note 6.
As the Rydberg-polariton density increased, the elastic collision rate also increased. Due to $\Omega_c =$ 1.0$\Gamma$, the values of $\Omega_p$ in the cases of Figs.~\ref{fig:Thermalization}(b) and \ref{fig:Thermalization}(c) well satisfied the perturbation condition. Thus, the propagation times of the probe light, i.e., the interaction times of the Rydberg polaritons, were approximately the same in the two cases. Under the same interaction time, the higher collision rate due to the larger Rydberg-polariton density produced a smaller width of the transverse momentum distribution or a lower effective transverse temperature, which is the expected outcome of the thermalization process. 

\FigSix

To check whether the uncertainty of $\delta_0$ could be significant on the probe beam profile due to the lensing effect, we measured the probe beam size not only at $\delta_0$ but also at $\delta_0$ $\pm$ 2$\pi\times$50~kHz. Figure~\ref{fig:beam_size} shows the measured beam widths at these two-photon detunings. As $\Omega_p = 0.1\Gamma$, $\delta_0$ = $-2\pi\times$85~kHz. As $\Omega_p = 0.2\Gamma$, $\delta_0$ = $-2\pi\times$180~kHz. Both values of $\delta_0$ were determined in the way illustrated by Fig.~\ref{fig:phi_versus_delta}. In Fig.~\ref{fig:beam_size}, the gray dashed line is the beam width measured without the presence of the atoms, and the black and red solid lines are the linear best fits of the experimental data. It is verified by the gentle slopes of the best fits that the uncertainty in $\delta_0$ and the lensing effect played insignificant roles in the measurement of the probe beam width. 

To verify that the observed reduction of the beam size on the EMCCD image is not caused by the nonlinear self-focusing effect of the probe beam, we also measured the probe beam size at $\Delta_c =$ $-1$$\Gamma$. As demonstrated by Fig.~\ref{fig:DDI}(d), the DDI-induced phase shift at $\Delta_c=-1\Gamma$ is 5$\sim$6 times smaller than that of at $\Delta_c =$ $+1$$\Gamma$. Consequently, the nonlinear self-focusing effect at $\Delta_c =$ $-1$$\Gamma$ is little as compared with that at $\Delta_c =$ $+1$$\Gamma$. Figure~\ref{fig:Thermalization_negative} shows the EMCCD images taken at $\Delta_c =$ $-1$$\Gamma$ with the same procedure and method as those in Fig.~\ref{fig:Thermalization}. The reduction of the probe beam size is still observed at $\Omega_p =$ 0.1$\Gamma$ (or 0.2$\Gamma$) by comparing Figs.~\ref{fig:Thermalization_negative}(a) and \ref{fig:Thermalization_negative}(b) [or  \ref{fig:Thermalization_negative}(c)] without and with the presence of atoms, respectively. Due to the large attenuation coefficient at $\Delta_c =$ $-1$$\Gamma$, the degree of the reduction in Fig.~\ref{fig:Thermalization_negative}(c) is not as significant as that in Fig.~\ref{fig:Thermalization}(c), but they are comparable. Thus, the reduction of the probe beam width cannot be explained by the nonlinear self-focusing effect, and it is the consequence of the transverse momentum distribution of the Rydberg polaritons being narrowed the thermalization process.

It is noteworthy that we did not observe any asymmetry in the images of Figs.~\ref{fig:Thermalization}(b) and \ref{fig:Thermalization}(c). In Ref.~\rcite{59}{Anisotropic_Interaction}, two Rydberg atoms were trapped in two microscopic dipole traps. The angle between the interatomic axis and the propagation direction of light fields was well defined. Thus, an anisotropic interaction as a function of the angle due to the Rydberg $D$ state was observed. In this work, the Rydberg atoms were randomly distributed. Angles between the interatomic axes and the propagation direction of light fields were also randomly distributed. Furthermore, we took the probe beam images or Rydberg-polariton momentum distributions in the plane transverse to the propagation direction of light fields or the quantization axis. As expected, the images of Figs.~\ref{fig:Thermalization}(b) and \ref{fig:Thermalization}(c) do not reveal any anisotropic interaction of the Rydberg $D$ state.

\FigSeven

\vspace*{\baselineskip}
\section{Discussion}

\indent Let us estimate the effective transverse temperature from Fig.~\ref{fig:Thermalization} to show the thermalization efficiency. Following the idea in Ref.~\rcite{28}{DSP_BEC_Fleischhauer2008}, the effective mass of Rydberg polaritons in the transverse direction is $m_\perp = \eta \hbar k_p \Gamma / \Omega_c^2$, where $\eta$ is the optical depth per unit length and k$_p$ is the wave vector of the probe light. Suppose the Rydberg polaritons obey the Maxwell-Boltzmann distribution. Then, the effective temperature in the transverse direction is given by
\begin{equation}
	T_{{\rm eff}\perp} = \hbar^2 \Delta_k^2/(2 m_\perp k_B),
\end{equation}
where $\Delta_k$ is the e$^{-1}$ half width of the momentum distribution and $k_B$ is the Boltzmann constant. Using the above formulas, we estimate the values of $T_{{\rm eff}\perp}$ corresponding to Figs.~\ref{fig:Thermalization}(d) and \ref{fig:Thermalization}(f) [or equivalently Figs.~\ref{fig:Thermalization}(a) and \ref{fig:Thermalization}(c)] to be about 3.1 and 1.2~$\mu$K, respectively. Thus the effective temperature $T_{{\rm eff}\perp}$ experiences a 2.6-fold reduction, as the Rydberg polaritons with the initial density of 2$\times$10$^9$~cm$^{-3}$ ($\Omega_p = 0.2$$\Gamma$) traverse the medium. Furthermore, Fig.~\ref{fig:Thermalization}(e) [or \ref{fig:Thermalization}(b)] for $\Omega_p = 0.1$$\Gamma$ indicates $T_{{\rm eff}\perp} =$ 2.0~$\mu$K. As $T_{{\rm eff}\perp}$ of Fig.~\ref{fig:Thermalization}(f) is compared with that of Fig.~\ref{fig:Thermalization}(e), a larger $\Omega_p$ corresponding to a higher polariton collision rate results in a better cooling effect. 

In addition to the elastic collision, the observed cooling effect can be assisted by the EIT bandwidth. After an elastic collision between two Rydberg polaritons, one polariton can increase the momentum and the other can decrease the momentum. The one with a larger momentum has a higher frequency of two-photon detuning and is more likely dissipated in the medium due to the EIT bandwidth, while the other with a smaller momentum has a lower frequency and can survive well. After many elastic collisions, the average energy or temperature of the Rydberg-polariton system becomes reduced.

The authors of Ref.~\rcite{28}{DSP_BEC_Fleischhauer2008} proposed to utilize stationary dark-state polaritons, i.e., stationary light based on the $\Lambda$-type EIT, to form the Bose-Einstein condensation (BEC). Please note that the interaction between such dark-state polaritons is induced by a far-detuned coupling scheme, e.g., the detuning used in Ref.~\rcite{28}{DSP_BEC_Fleischhauer2008} is 50 times exceeding the spontaneous decay rate. Thus, the scattering cross section of dark-state polaritons is far smaller than that of Rydberg polaritons interacting due to the strong interaction between the Rydberg atoms. Once Rydberg polaritons are made stationary, the BEC formulism in Ref.~\rcite{28}{DSP_BEC_Fleischhauer2008} can be readily applied to the Rydberg-polariton system. The critical temperature $T_\text{c}$ for the polariton BEC under the cylindrical symmetry is given by\ncite{DSP_BEC_Fleischhauer2008}
\begin{eqnarray}
	T_{\rm c} &=& \frac{\pi \hbar^2 \sqrt[3]{n_p^2}}{k_B \sqrt[3]{m_\perp^2 m_\parallel}}, \\
	m_\perp &=& \frac{\hbar \eta k_p \Gamma}{\Omega_c^2}, \\
	m_\parallel &=& \frac{\hbar \eta^2 \Gamma^2}{8 \Omega_c^2 \Delta_c}, 
\end{eqnarray}
where $n_p$ is the Rydberg-polariton density, $m_\perp$ and $m_\parallel$ are the transverse and longitudinal effective masses, $\eta$ is the optical depth per unit length, and $k_p$ is the wave vector of the probe light. However, in the present study we dealt with a two-dimensional Rydberg-polariton system. The Rydberg polaritons propagated along the longitudinal direction. The average longitudinal kinetic energy of each Rydberg polariton ($= \hbar k_p v_g / k_B$ where $v_g$ is the group velocity) given by 140 mK was much larger than the transverse temperature. To achieve BEC, we need to make Rydberg polaritons stationary, and further enhance OD to have a sufficient interaction time, i.e., life time of the stationary light\cite{SLP_OurPRL2012}. It is also desirable to add an artificial trap to enable an efficient evaporative cooling or energy dissipation. Using $\eta = 160$ cm$^{-1}$, $\Omega_c =1.0$$\Gamma$, and $\Delta_c =1.0$$\Gamma$, we estimate $T_{\rm c} =$ 2.0 mK for the stationary Rydberg polaritons of $n_p =$ 2$\times$10$^9$~cm$^{-3}$.

Up to now, the BECs of various kinds of polaritons have been realized in cavity systems\ncite{EPBEC_nature2006, EPBEC_science2007, PhotonBEC_nature2010, EPBEC_RMP2010}. This work deals with the cavity-free high-OD medium, in which the interaction time between the Rydberg polaritons is analogous to the storage time of polaritons in a cavity with a $Q$ factor greater than $10^9$. While the current systems of polariton BECs are all two-dimensional, a three-dimension BEC can be formed in the system of Rydberg or dark-state polaritons\ncite{DSP_BEC_Fleischhauer2008}. In addition, as compared with other kinds of polaritons, the condensate of Rydberg polaritons can have a much longer life time due to a low decoherence rate in the Rydberg-EIT system, offering more opportunities for many-body physics of polariton systems.

\vspace*{\baselineskip}
\section{Method}

\vspace*{0.5\baselineskip}
\subsection{Laser fields preparation.}
 The probe and coupling fields were generated by a homemade diode laser and a blue laser system (Toptica TA-SHG pro), respectively. The frequency stabilizations of the probe and coupling lasers are described as follows. We utilized the injection lock scheme to stabilize the frequency of the probe laser. In the scheme, the probe laser was used as the slave, and an external-cavity diode laser (ECDL) of Toptica DLC DL pro with the wavelength of about 780~nm was employed as the master. We utilized the Pound-Drever-Hall scheme and the saturated absorption spectroscopy to lock the frequency of the ECDL with a heated vapor cell of Rb atoms. The blue laser had the wavelength of about 482~nm. Furthermore, we used the Pound-Drever-Hall scheme and the EIT spectrum to lock the frequency of the blue laser with another heated vapor cell. The ECDL and blue laser fields counter-propagated in the measurement. We were able to lock the sum of the two laser frequencies to the EIT transition with the root-mean-square fluctuation of around 150~kHz\ncite{REIT_OurPRA2019}.

\vspace*{0.5\baselineskip}
\subsection{Determination of experimental parameters.}
 The parameters of the OD ($\alpha$), coupling Rabi frequency ($\Omega_c$), and intrinsic decoherence rate ($\gamma_0$) were determined experimentally with the same method used in Ref.~\rcite{47}{REIT_OurPRA2019}. Details of the determination procedure can be found in Supplementary Note 2. We set $\Omega_c = 1.0\Gamma$ and were able to maintain $\gamma_0$ around 0.012(1)$\Gamma$ throughout the experiment. This $\gamma_0$ included the effects of laser frequency fluctuation, Doppler shift, and other decoherence processes that appear in the $\Lambda$-type EIT system\ncite{REIT_OurPRA2019}. 

\vspace*{0.5\baselineskip}
\subsection{Beat-note interferometer.}
 We employed the beat-note interferometer to measure the phase shift of the probe field induced by the atoms. The concept and illustration of the beat-note interferometer can be found in Refs.~\rcite{60}{beatnote} and \rcite{61}{phasevariation}. In the absence of the dipole-dipole interaction (DDI), the phase shift of the probe field also enabled us to precisely determine the probe frequency for the zero two-photon detuning, i.e., $\delta = 0$, at a given coupling detuning, $\Delta_c$. This is because the phase shift is equal to $\tau_d \delta$ according to the non-DDI EIT theory, where $\tau_d$ is the propagation delay time. The data of attenuation coefficients and phase shifts under the DDI effect presented in Fig.~\ref{fig:DDI} were taken at $\delta = 0$. See Supplementary Note 1 for more details. 

\vspace*{\baselineskip}
{\noindent \bf \small Data availability} \\
{\footnotesize \noindent
The data and information within this paper are available from the corresponding author upon request.
}

\vspace*{\baselineskip}

\vspace*{\baselineskip}
{\noindent \bf \small Acknowledgments} \\
{\footnotesize \noindent
This work was supported by Grant Nos.~105-2923-M-007-002-MY3, 108-2639-M-007-001-ASP, and 109-2639-M-007-002 of the Ministry of Science and Technology of Taiwan, Project No.~TAP LLT-2/2016 of the Research Council of Lithuania, and Project No.~LV-LT-TW/2018/7 of the Ministry of Education and Science of Latvia. JR and GJ also acknowledge a support by the National Center for Theoretical Sciences, Taiwan.}

\vspace*{\baselineskip}
{\noindent \bf \small Author contributions} \\
{\footnotesize \noindent
I.A.Y. conceived the idea of weakly-interacting many-body system of Rydberg polaritons. S.-S.H., G.J., J.R., T.K., M.A., and I.A.Y. developed the theoretical foundation for the idea. B.K., K.-T.C., Y.-C.C., Y.-F.C. and I.A.Y. designed the experimental setup and methods. B.K., K.-T.C., S.-Y.W., and K.-B.L. built the setup, carried out the experiment, and analyzed the data. The manuscript was written by B.K., S.-S.H., and I.A.Y. with helps from all the other authors.}

\vspace*{\baselineskip}
{\noindent \bf \small Competing interest}\\
{\footnotesize \noindent
The authors declare no competing interests.
}

\vspace*{\baselineskip}
{\noindent \bf \small Additional information} \\
{\footnotesize \noindent
{\bf Supplementary information} is available for this paper on the website.\\
{\bf Correspondence} and requests for materials should be addressed to I. A.Y. 
}


\end{document}


\title{
Supplementary Information\\ \vspace*{\baselineskip}
A weakly-interacting many-body system of Rydberg polaritons based on electromagnetically induced transparency
}

\author{Bongjune Kim$^1$}
\author{Ko-Tang Chen$^1$} 
\author{Shih-Si Hsiao$^1$} 
\author{Sheng-Yang Wang$^1$}
\author{Kai-Bo Li$^1$}
\author{Julius Ruseckas$^2$}
\author{Gediminas Juzeli\={u}nas$^2$}
\author{Teodora Kirova$^3$}
\author{Marcis Auzinsh$^4$}
\author{Ying-Cheng Chen$^{5,7}$}
\author{Yong-Fan Chen$^{6,7}$}
\author{Ite A. Yu$^{1,7,*}$}

\affiliation{
$^{1}$Department of Physics, National Tsing Hua University, Hsinchu 30013, Taiwan \\
$^{2}$Institute of Theoretical Physics and Astronomy, Vilnius University, Saul\.{e}tekio 3, 10257 Vilnius, Lithuania \\
$^{3}$Institute of Atomic Physics and Spectroscopy, University of Latvia, LV-1586 Riga, Latvia \\
$^{4}$Laser Centre, University of Latvia, LV-1002, Riga, Latvia \\
$^{5}$Institute of Atomic and Molecular Sciences, Academia Sinica, Taipei 10617, Taiwan \\
$^{6}$Department of Physics, National Cheng Kung University, Tainan 70101, Taiwan \\
$^{7}$Center for Quantum Technology, Hsinchu 30013, Taiwan \\
\hspace*{2cm}
$^{*}$email: yu@phys.nthu.edu.tw 
\hspace*{2cm}
}



\maketitle
\vspace*{-3\baselineskip}


\newcommand{\FigSOne}{
	\begin{figure}[b]
	\center{\includegraphics[width=\textwidth]{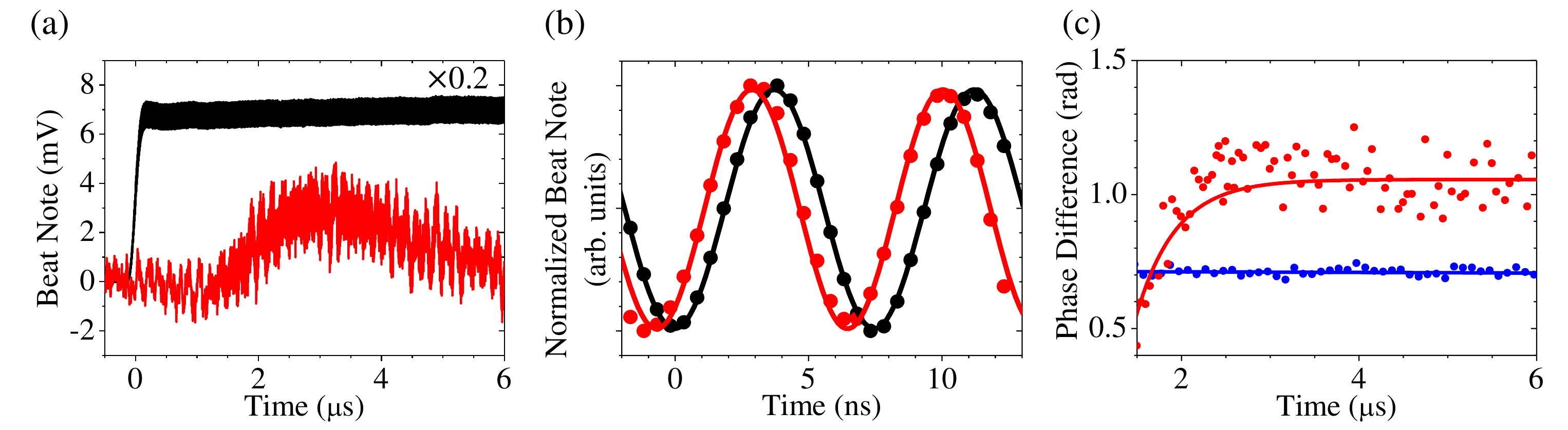}}
	\caption{Determination of the phase shift with the beat-note interferometer. (a) Representative data of the probe and reference beat notes as functions of time are shown by red and black lines. The high-frequency component of the lines is the beat note. The black line is scaled down by a factor of 0.2. (b) We zoom in the data around 5.2~$\mu$s of (a) to demonstrate the phase difference between probe (red circles) and the reference (black circles) beat notes. Red and black lines are the best fits of sinusoidal functions. (c) Phase difference measured with the atoms (red circles) and that measured without the atoms (blue circles) as functions of time. The two measurements were taken consecutively. Red line is the exponential best fit of the red circles and blue line is the average value of the blue circles. The experimental data were measured at $\alpha$ (optical depth) = 80, $\Omega_c$ (coupling Rabi frequency) $= 1.0\Gamma$, $\Delta_c$ (coupling detuning) $= 0$, $\delta$ (two-photon detuning) $= 0$, and $\Omega_p$ (probe Rabi frequency) $= 0.125\Gamma$, where $\Gamma =$ 2$\pi$$\times$6~MHz in this work. 
}
	\label{fig:phase}
	\end{figure}
}
\newcommand{\FigSTwo}{
	\begin{figure}[b]
	\center{\includegraphics[width=0.35\columnwidth]{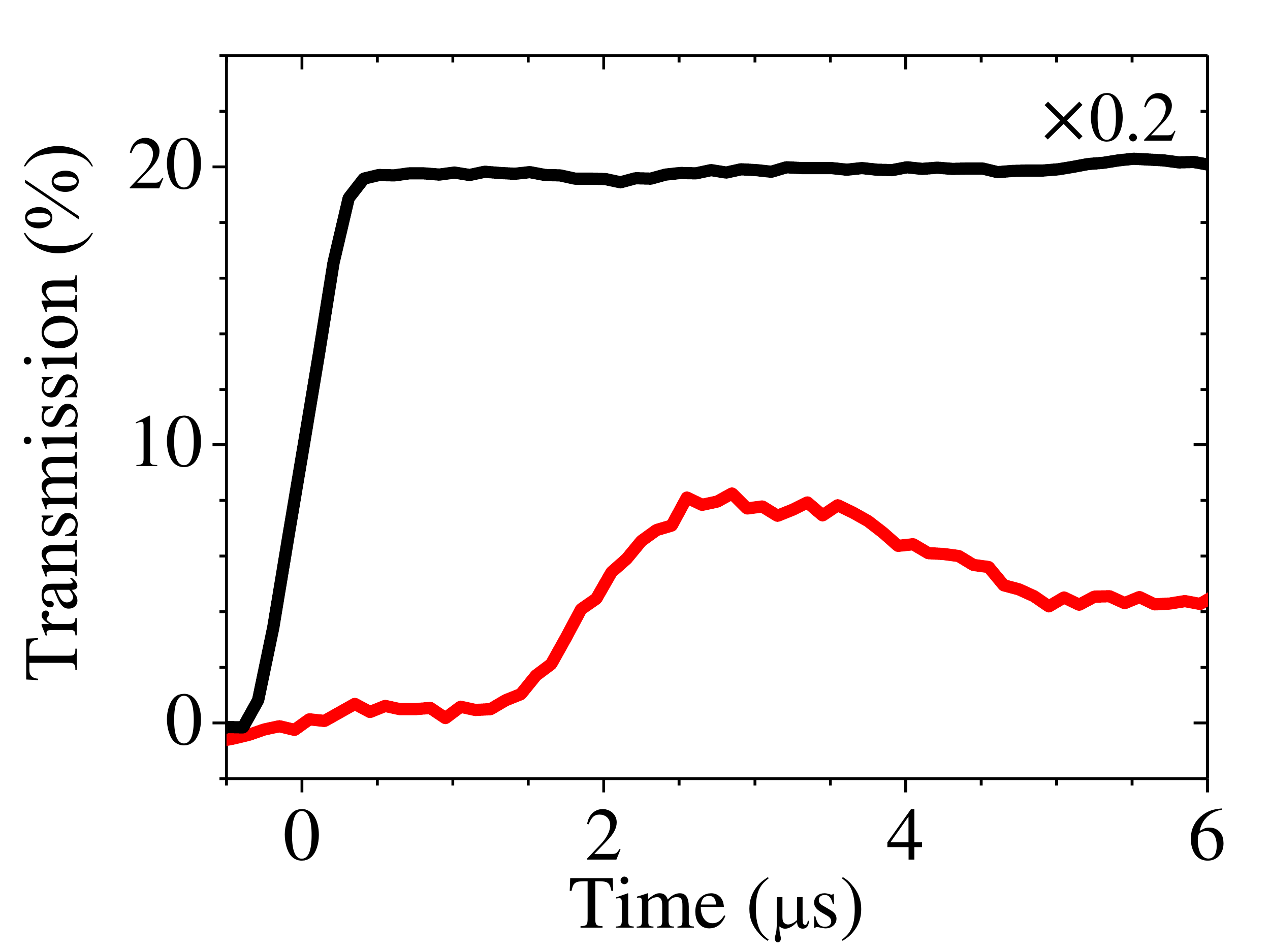}}
	\caption{Representative data of the probe transmission as a function of time taken at $\alpha$ (optical depth) = 80, $\Omega_c = 1.0\Gamma$, $\Delta_c = 0$, $\delta = 0$, and $\Omega_p = 0.125\Gamma$. Black and red lines are the input and output probe signals. The black line is scaled down by a factor of 0.2. We took the mean output transmission in the time interval between 5 and 6~$\mu$s to calculate the attenuation coefficient.
}
	\label{fig:beta}
	\end{figure}
}
\newcommand{\FigSThree}{
	\begin{figure}[b]
	\center{\includegraphics[width=0.35\columnwidth]{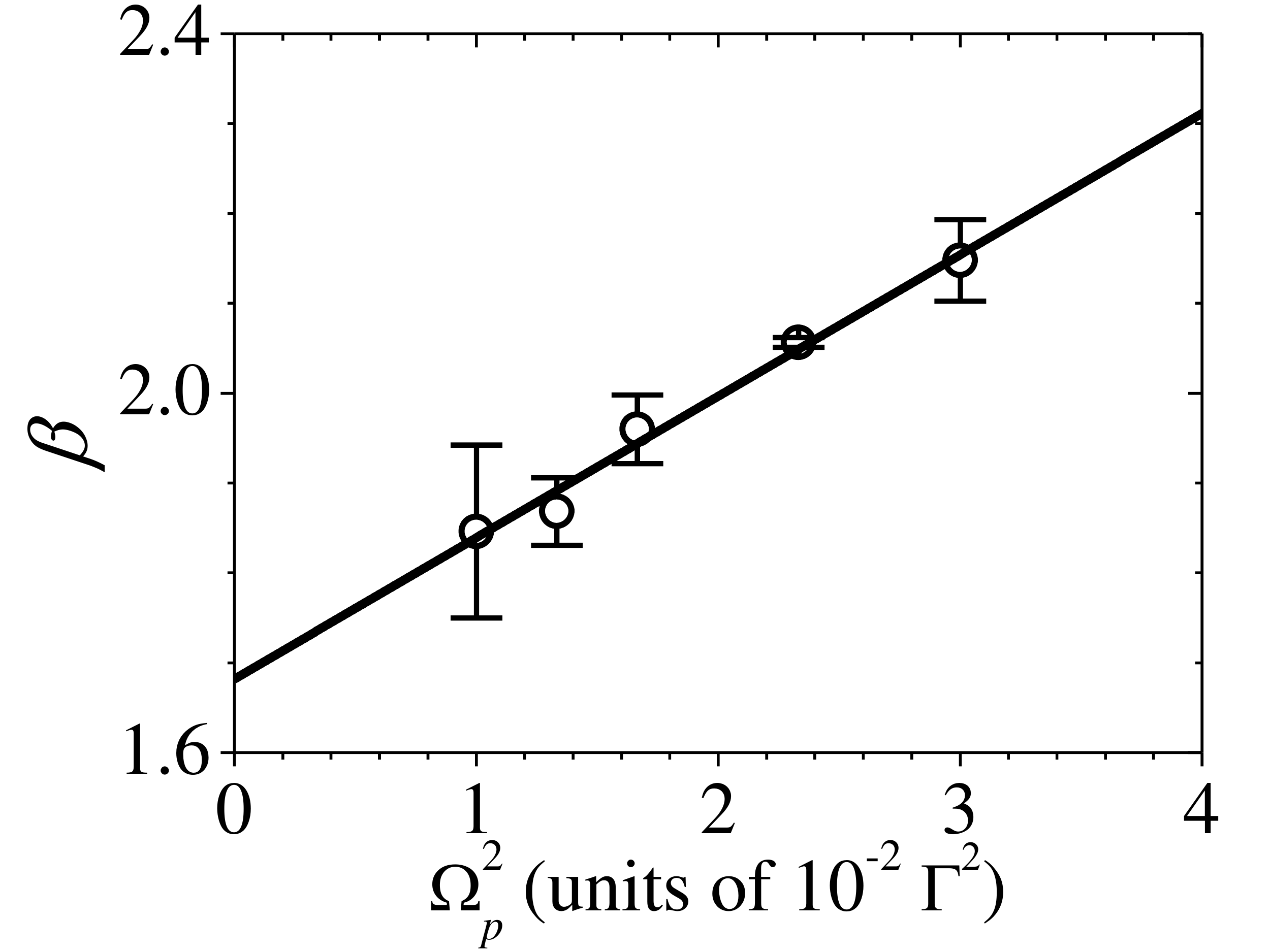}}
	\caption{Attenuation coefficient, $\beta$, as a function of $\Omega_p^2$ in the case II at $\alpha$ (optical depth) = 82, $\Omega_c =$ 1.0$\Gamma$, $\Delta_c =$ 1.0$\Gamma$, and $\delta = 0$. Circles are the experimental data and straight line is the best fit. The slope, $\chi_{\beta}$, of the best fit is (16$\pm$1)$/\Gamma^2$ here, while $\chi_{\beta}$ is (17$\pm$4)$/\Gamma^2$ in the case I, measured at $\alpha$ = 81, $\Omega_c =$ 1.0$\Gamma$, $\Delta_c =$ 1.0$\Gamma$, and $\delta = 0$. The error bars represent the standard deviation of measured values.
}
	\label{fig:new_path_beta}
	\end{figure}
}
\newcommand{\FigSFour}{
	\begin{figure}[b]
	\center{\includegraphics[width=0.35\columnwidth]{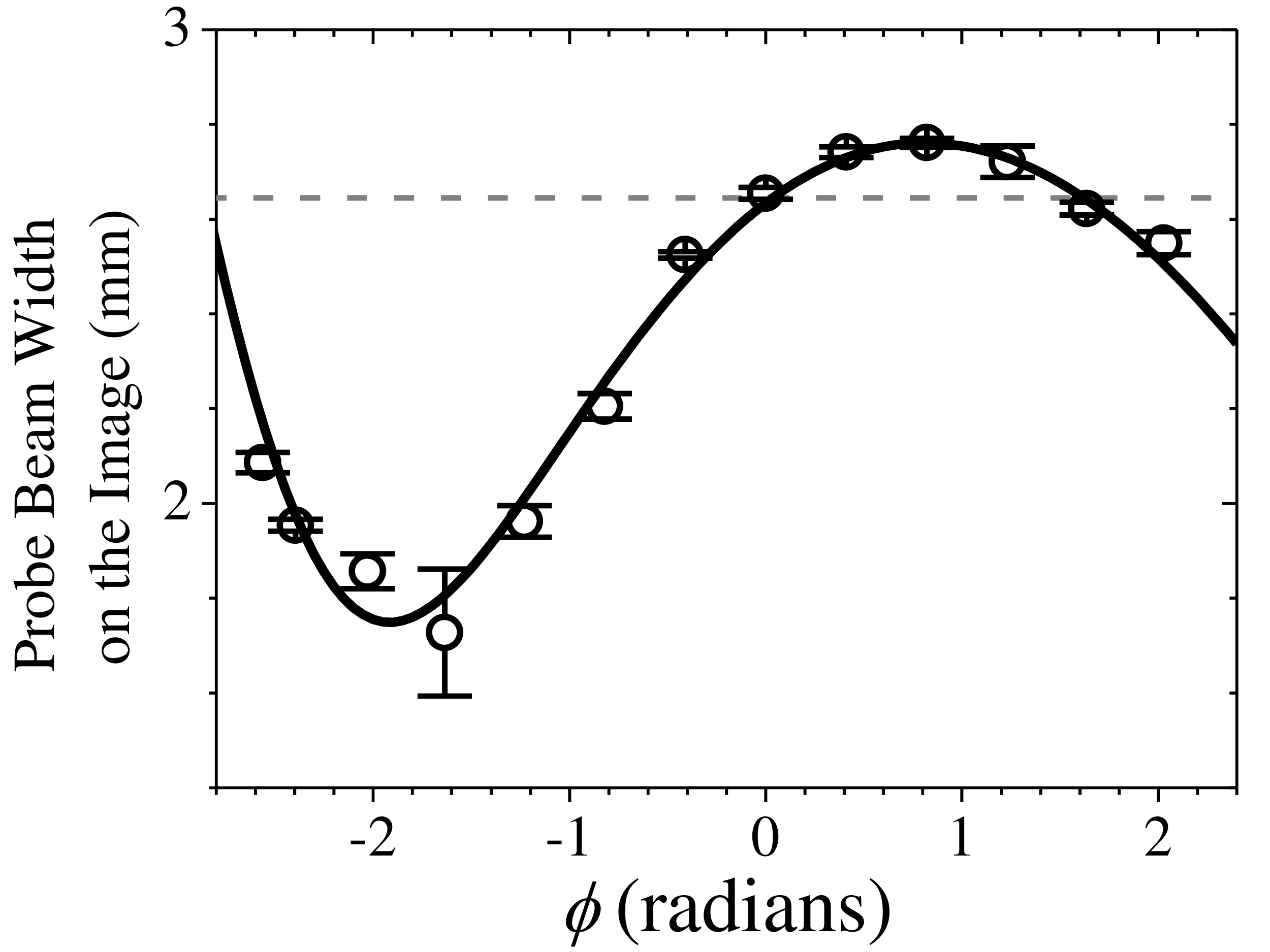}}
	\caption{Demonstration of the lensing effect. We measured the e$^{-1}$ full width of the probe beam profile on the EMCCD image in the case II as a function of the probe phase shift. Circles are the experimental data taken at $\alpha$ (optical depth) = 13, $\Omega_c =$ 1.0$\Gamma$, $\Delta_c = 0$, and $\Omega_p =$ 0.15$\Gamma$. Black line is the best fit calculated with the consideration of the atoms behaving like a GRIN lens. The best fit gives $d =$ 19.5~mm, $p =$ 300~rad$\cdot$mm$^2$, and $z_0$ = 1.8~mm, where the meanings of $d$, $p$, and $z_0$ can be found in the text. Gray dashed line is the width measured without the presence of the atoms. The error bars represent the standard deviation of measured values.
}
	\label{fig:lensing_effect}
	\end{figure}
}
\newcommand{\FigSFive}{ 
	\begin{figure}[t]
	\center{\includegraphics[width=0.35\columnwidth]{_FigS5}}
	\caption{Verification of the uncertainty in the two-photon detuning as well as the lensing effect being insignificant in the measurement of the probe beam width, i.e., the transverse momentum distribution of the Rydberg polaritons. Black and red circles are the experimental data of the e$^{-1}$ full width of the output probe beam versus the two-photon detuning. The black (or red) data points were taken at OD = 82, $\Omega_c =$ 1.0$\Gamma$, $\Delta_c = 0$, and $\Omega_p =$ 0.10$\Gamma$ (or 0.20$\Gamma$). Black and red lines are the linear best fits of the experimental data, and their slopes are 2.5~$\upmu$m/kHz and 2.1~$\upmu$m/kHz, respectively. Gray dashed line indicates the probe beam width measured without the presence of the atoms, i.e., the input probe beam width.
}
	\label{fig:beam_size}
	\end{figure}
}

\section{Beat-note interferometer for the phase measurement}
\label{sec:BNI}
\FigSOne

As shown by the brown path in \MAIN{Fig.~1(b) of the main text}, the zeroth-order beam of the acousto-optic modulator (AOM) was unblocked when we performed the phase measurement. The first-order beam of the AOM was the probe field. We combined the probe beam and the zeroth-order beam with a 50/50 beam splitter, BS$_{a}$, in the figure to form the beat note, which had a fixed and stable frequency set by the AOM's driving frequency. The polarization-maintained optical fiber (PMF) can ensure both beams to completely overlap and have the same polarization. Coming out of the PMF, the beat-note signal was split into two by another 50/50 beam splitter, BS$_{b}$. One, named the reference beat note, did not pass through the atoms and was detected by a photo detector (PD). The other, name the probe beat note, passed through the atoms and was detected by the PMT. Both beat notes were measured simultaneously. We compared the photomultiplier tube (PMT) and PD output signals to determine the phase difference between the probe and reference beat notes. Since the beat-note frequency (or wavelength) was around 133~MHz (or 2.3 m), the phase difference measured by the beat note interferometer was insensitive to the change or flucutation of the optical path \cite{beatnote}. The insensitivity was demonstrated by the result that the phase difference measured without the atoms was a stable constant. In addition, the frequency of the zeroth-order beam was far away from the transition frequency. Thus, the measured phase difference was mainly contributed from the phase shift of the probe field. We subtracted the phase difference measured with the atoms from that measured without the atoms to obtain the phase shift of the probe field induced by the atoms.

We carefully set the two-photon detuning to zero at a given coupling detuning ($\Delta_c$), before measuring the additional phase shift and attenuation induced by the DDI. To determine the probe frequency for $\delta$ (two-photon detuning) $=0$, the experimental condition was changed to $\alpha$ (optical depth abbreviated as OD) = 22$\sim$26 and $\Omega_p \approx$ 0.05$\Gamma$ such that the dipole-dipole interaction (DDI) effect was negligible, where $\Gamma$ = 2$\pi\times$6~MHz is the spontaneous decay rate of the excited state of the probe transition. During the beat-note measurement, the probe beam was a square pulse and the zeroth-order beam was a continuous wave. The zeroth-order beam was far detuned by about $-22$$\Gamma$ and had the weak Rabi frequency of around 0.2$\Gamma$. Thus, the delay times of the probe field (as well as the ODs and decoherence rates of the experimental system) with and without the zeroth-order beam were nearly the same. Beat-note signals were acquired at a speed of 2$\times 10^9$ samples per second and stored in the segmented memory by the oscilloscope. We averaged beat-note data for 936 times and determined the phase difference between the probe and reference beat notes with the sinusoidal best fits of the data. For example, the high-frequency components of the lines in Supplementary Fig.~\ref{fig:phase}(a) and the circles in Supplementary Fig.~\ref{fig:phase}(b) demonstrate the beat-note data, while the red and black lines in Supplementary Fig.~\ref{fig:phase}(b) are the best fits. Once obtaining the probe frequency for $\delta = 0$, we immediately switched back to the experimental condition of $\alpha$ (OD) $\approx$ 81 and $\Omega_p \geq 0.1\Gamma$, and took the data of the attenuation coefficient and phase shift under the DDI effect as shown by \MAIN{Fig.~2 of the main text}.

\section{Determination of experimental parameters}
\label{sec:parameter}

The experimental parameters of the coupling Rabi frequency, $\Omega_c$, the optical depth (OD), $\alpha$, and the intrinsic decoherence rate, $\gamma_0$, were determined in the same methods described in Ref.~\cite{REIT_OurPRA2019}. As the examples of the determination methods, Figs.~7(a)-7(c) of this reference demonstrated the data of a similar Rydberg-EIT experiment.

We determined $\Omega_c$ by the frequency separation of two minima, i.e. the Autler-Towns splitting, in the probe transmission spectrum measured at the coupling detuning, $\Delta_c$, of nearly zero and OD $\approx 1$. An example of the spectrum is shown in Fig.~7(a) of Ref.~\cite{REIT_OurPRA2019}. Suppose the frequencies of the two minima are denoted as $\delta_-$ and $\delta_+$. Then, $\Omega_c$ was obtained by the relation of 
\begin{equation}
\label{eq:AT-splitting}
	\delta_{+} - \delta_{-} \approx \Omega_c + \frac{\Delta_c^{2}}{2\Omega_c} 
\end{equation}
under the condition of $\Delta_c \ll \Omega_c$. The difference between the magnitudes of $\delta_-$ and $\delta_+$ is given by 

\begin{equation}
	|\delta_+| - |\delta_-| = \Delta_c.
\end{equation}
Using the above difference, we carefully minimized $|\Delta_c|$ in the measurement such that the correction term $\Delta_c^2/2\Omega_c$ in Eq.~(\ref{eq:AT-splitting}) was about $1.3\times 10^{-4}\Gamma$. The condition of $\Delta_c = 0$ was also determined in this step. To obtain the Autler-Towns splitting, we swept the probe frequency by varying the driving frequency of the AOM shown in \MAIN{Fig.~1(b) of the main text}. The double-pass scheme of the AOM made the input probe power relatively stable during each run of the frequency sweeping. To prevent the spectrum from being distorted by the transient effect, a sufficiently slow sweeping rate of 240 kHz/$\mu$s was utilized.

After the value of $\Omega_c$ was known, we determined the OD, i.e., $\alpha$, in a designated study by measuring the propagation delay time of the probe pulse. An example of the delay time measurement is shown in Fig.~7(b) of Ref.~\cite{REIT_OurPRA2019}. A short Gaussian input pulse with e$^{-1}$ full width of 0.66~$\mu$s was used in the measurement. To minimize the DDI effect, we employed a weak probe pulse with the peak Rabi frequency of 0.05$\Gamma$. According to the non-DDI EIT theory, the delay time, $\tau_d$ is given by 
\begin{equation}
	\tau_d = \frac{\alpha \Gamma}{\Omega_c^2}.
\end{equation}
We used the measured value of $\tau_d$ to obtain the OD of the experiment in the designated study. The experimental data in the figures of the main text were taken at the OD around 81, which was determined by the results of $\tau_d =$ 2.1~$\mu$s at $\Omega_c =$ 1.0$\Gamma$ and $\tau_d =$ 1.0~$\mu$s at $\Omega_c =$ 1.5$\Gamma$ in the measurement of slow light.

The intrinsic decoherence rate $\gamma_0$ was determined by measuring the probe transmission at both of the two-photon and one-photon resonances, i.e., not only $\delta = 0$ but also $\Delta_c = 0$. In the measurement, we set $\Omega_c =$ 1.0$\Gamma$ and $\alpha = 24$. The input Gaussian pulse of the probe field had the e$^{-1}$ full width of 7~$\mu$s and the peak Rabi frequency of 0.05$\Gamma$. The low OD and the weak probe pulse were used to made the DDI effect negligible. The long probe pulse made the measured transmission as the steady-state result. According to the non-DDI EIT theory, the peak transmission $T_{\rm max}$, i.e., the transmission at $\delta = 0$, is given by
\begin{equation}
	T_{\rm max} = \exp \left( -\frac{2 \alpha \Gamma}{\Omega_c^2} \gamma_0 \right).
\end{equation}
The measured value of $T_{\rm max}$ inferred the intrinsic decoherence rate of the experiment. We determined $\gamma_0$ before each study and its values maintained at (11$\sim$13)$\times$$10^{-3}$$\Gamma$ throughout this work. 

\section{Measurements of transmission and phase shift of the output probe field}

In this section, we will describe the measurement procedure of the probe transmission and phase shift under the DDI effect. To make the DDI effect significant, a high OD of 81$\pm3$ and a moderate value of $\Omega_c$ of (1.0$\pm$0.03)$\Gamma$ were used and kept constant in all of the DDI studies presented in the paper. The input probe field was a long square pulse with 0.1$\Gamma$ $\leq \Omega_p \leq$ 0.2$\Gamma$. As compared with $\Omega_c$, such weak $\Omega_p$ makes the probe field as the perturbation. Under the perturbation limit and without any DDI effect, the attenuation coefficient and phase shift of the output probe field do not change by varying $\Omega_p^2$.

\MAIN{Figures~2(a) and 2(c) of the main text} show the attenuation coefficient and phase shift of the output probe field as functions of $\Omega_p^2$. There are five different values of the coupling detuning, $\Delta_c$, in each figure. We performed a set of consecutive measurements at each value of $\Delta_c$. In each measurement set, we carried out the following steps: (i) The experimental parameters of $\Omega_c$, $\alpha$, and $\gamma_0$ were verified by the methods described in \ref{sec:parameter}. (ii) We set and checked a designated value of $\Delta_c$. (iii) The probe frequency for the two-photon resonance condition, i.e., $\delta = 0$, was searched for and determined by the beat-note interferometer illustrated in \ref{sec:BNI}. (iv) We measured the probe transmissions or phase shifts at different values of $\Omega_p^2$ consecutively. (v) The steps (i), (ii), and (iii) were performed in the reverse order to ensure that the experimental condition did not change. Since it took about one hour to complete the steps (iii) and (iv), we confirmed the two-photon frequency had an uncertainty of merely $\pm$20~kHz due to the long-term drift. Because of the above steps, it can be certain that the additional attenuation and phase shift were contributed mainly from the variation of $\Omega_p^2$, i.e. from the DDI effect, but very little from the fluctuation or change of $\delta$ and $\gamma_0$.

\FigSTwo

The representative data of the input and output probe transmissions are shown in Supplementary Fig.~\ref{fig:beta}. The data were averaged 512 times by the oscilloscope. We calculated the logarithm of the mean output transmission in the time interval between 5 and 6~$\mu$s to obtain the value of $-\beta$, where $\beta$ is the attenuation coefficient. The attenuation coefficients at different values of $\Omega_p^2$ and $\Delta_c$ shown in \MAIN{Fig.~2(a) of the main text} were determined in the same way. The representative data of the reference and probe beat notes are shown in Supplementary Fig.~\ref{fig:phase}(a) and illustrated in \ref{sec:BNI}. The data look like some high-frequency components added to the signals shown in Supplementary Fig.~\ref{fig:beta}. These high-frequency components are the beat notes. We zoomed in each time interval of 50 ns, and determined the phase difference between the probe and reference beat notes. In the absence of the atoms, the phase difference was a stable constant of 0.71 radians as shown by the blue circles in Supplementary Fig.~\ref{fig:phase}(c). In the presence of the atoms, the representative data of phase difference as a function of time are shown by the red circles in Supplementary Fig.~\ref{fig:phase}(c). The red line in the figure is the best fit of the red circles. We subtracted the steady-state value of the red line from 0.71 radians to obtain the phase shift of the probe field induced by the atoms. The phase shifts at different values of $\Omega_p^2$ and $\Delta_c$ shown in \MAIN{Fig.~2(c) of the main text} were determined in the same way.

\section{Measurement of transverse momentum distribution of Rydberg polaritons}
\label{sec:thermalization}

To observe the evidence of the DDI-induced thermalization process in the Rydberg-polariton system, we studied the transverse momentum distribution of the Rydberg polaritons by taking images of the probe beam. After entering the medium of the atoms, the transverse momentums of the input probe photons were converted to those of the Rydberg polaritons. In the medium, the DDI can induce the elastic collisions among the Rydberg polaritons and reduce the width of their transverse momentum distribution. As the probe beam was exiting the medium, the transverse momentums of the Rydberg polaritons were converted back to those of the output photons. The images of the input and output probe beam profiles revealed the initial and final momentum distributions of the Rydberg polaritons. It can be seen as the consequence of cooling after the thermalization process that the output probe beam size becomes smaller than the input one, i.e., the width of the final transverse momentum distribution becomes narrower than that of the initial one. However, the lensing effect of the atoms can also change the probe beam size. The condition of the zero phase shift, i.e. $\phi = 0$ can avoid the lensing effect, which will be explained in \ref{sec:lensing_effect}. For each of different values of $\Omega_p^2$, we first searched for the two-photon detuning, $\delta_0$, corresponding to $\phi = 0$, and then took the image of the output probe beam profile at $\delta_0$.  

To clearly observe the thermalization effect, we made the Rydberg polaritons have a large initial transverse momentum distribution. This was achieved by reducing the waist of the input probe beam by adding the lens pair of L4 and L5 shown in \MAIN{Fig.~3 of the main text}. With (and without) the lens pair, the e$^{-1}$ full width of the probe beam at the position of atom cloud center was 39~$\mu$m (and 130~$\mu$m). As shown in \MAIN{Fig.~1(b) of the main text}, the moveable mirror M$_b$ directed the output probe beam to the electron-multiplying charge-coupled device (EMCCD) camera instead of to the PMT. Let's call the experimental system without (and with) the installation of L4, L5, and M$_b$ as case I (and case II). The experimental parameters of $\Omega_c$ and OD of the two cases under the same settings differed a little, i.e., less than 2\%. We tuned the settings slightly to make the measured $\Omega_c$ and OD of the two cases become the same. Then, the measured decoherence rates, $\gamma_0$, had no observable difference. It took about a minute to switch from one case to another. We switched back and forth between the two cases in  some measurements, and the experimental parameters of each case under the same settings were unchanged throughout the day. 

\FigSThree

Using nearly the same $\Omega_c$, OD, and $\gamma_0$ in the measurements of the two cases, we also confirmed that the slope, $\chi_{\beta}$, of the attenuation coefficient, $\beta$, versus $\Omega_p^2$ in the case I was consistent with that in the case II. Supplementary Fig.~\ref{fig:new_path_beta} shows $\beta$ as a function of $\Omega_p^2$ which was measured in the case II at the zero two-photon detuning, i.e. $\delta = 0$. In the measurement, the input probe field was a Gaussian pulse with the e$^{-1}$ full width of 12~$\mu$s, since we took the EMCCD image with the similar input pulse. Derived from the slope of the straight line in Supplementary Fig.~\ref{fig:new_path_beta}, the value of $\chi_{\beta}$ was (16$\pm$1)/$\Gamma^2$. According to the data in \MAIN{Fig.~2 of the main text}, $\chi_{\beta}$ under nearly the same experimental parameters in the case I was (17$\pm$4)/$\Gamma^2$. Considering the uncertainties, the difference between the two  $\chi_{\beta}$ values is acceptable.

In the case II, the probe beam size was significantly reduced. Even with the Rabi frequency of 0.2$\Gamma$, the probe power was still too weak to produce reasonable beat-note signals. Therefore, the beat-note interferometer was not able to provide a sufficient accuracy for the determination of the probe phase shift in the case II. Instead, we searched for the two-photon detuning, $\delta_0$ corresponding to the zero phase shift, i.e, $\phi = 0$, in the case I. Since the values of $S_{\rm DDI}$ of the two cases were consistent, $\delta_0$ found in the case I can be applied to the case II.

The procedure of taking the image of the output probe beam profile is summarized in the following steps: (i) The experimental parameters were determined and set to the designed values. (ii) The probe frequency for $\delta = 0$ was determined in the case I by the method described in \ref{sec:BNI}. (iii) At the above experimental parameters, we measured $S_{\rm DDI}$ in the case II with the method similar to that shown in Supplementary Fig.~\ref{fig:new_path_beta}, and confirmed that the measured value was consistent with the value of $S_{\rm DDI}$ in the case I. (iv)  The two-photon detuning, $\delta_0$, corresponding to $\phi=0$ was determined in the case I. Representation data are shown in \MAIN{Fig.~4 of the main text}. We fitted the data with a straight line, and the best fit determined $\delta_0$. The uncertainty of $\delta_0$ was $\pm$2$\pi\times$30~kHz. (v) At $\delta_0$, we took images of the output probe beam profile in the case II. We repeated the steps (iv) and (v) for different values of $\Omega_p$. (vi) The value of $S_{\rm DDI}$ in the case II was measured again to verify it was unchanged during the steps (iv) and (v).

\section{Study on the lensing effect} 
\label{sec:lensing_effect}

In this section, we report the study on the lensing effect of the atoms. We experimentally demonstrated that condition of the zero phase shift, i.e., $\phi =0$, can avoid the lensing effect. Images of the output probe beam profiles were taken by the EMCCD for the study. Since the beam profile can also be affected by the thermalization process induced by the elastic collisions between the Rydberg polaritons, we performed the study with a low OD of 13 to minimize the DDI effect.

\FigSFour

Supplementary Fig.~\ref{fig:lensing_effect} shows the e$^{-1}$ full width of the output probe beam profile as a function of the phase shift. The width was measured by the EMCCD image in the case II. The phase shift was converted from the two-photon detuning, $\delta$, used in the measurement with the following formula:
\begin{equation}
	\phi = \frac{\alpha\Gamma}{\Omega_c^2} \delta.
\end{equation}
We verified the above formula and determined the probe frequency corresponding to $\delta$ with the beat-note interferometer in the case I. When taking images, we employed the Gaussian probe pulse of the e$^{-1}$ full width of 12~$\mu$s. To produce the images of a better signal-to-noise ratio, the peak Rabi frequency of the pulse was 0.15$\Gamma$.

The gray dashed line in Supplementary Fig.~\ref{fig:lensing_effect} represents the beam width measured without the presence of the atoms. It can be clearly seen that the experimental data point of $\phi = 0$ nearly locates on the gray dashed line, indicating that the condition of $\phi = 0$ can avoid the lensing effect of the atoms. Such a result is also expected from the theory as shown by the black line in the figure. To calculate the black line, we utilized the ray-tracing method for Gaussian beams. In addition, we considered that the atoms behaved like a gradient index (GRIN) lens of which the refractive index, $n$, changes with the transverse distance, $r$, from the symmetric or central axis in the direction of light propagation. The ABCD matrix of a GRIN lens is given by
\begin{equation}
	\MT{\cos (gd)}{\sin (gd)/g}{-g\sin (gd)}{\cos (gd)},
\end{equation}
where $d$ is the thickness of the lens and $g$ is defined by $n(r) = n_0 - g^2 r^2/2$ where $n_0$ the refractive index along the central axis \cite{GRINlens1990}. 


Since the refractive index is proportional to the phase shift, $\phi$, of the atoms, the following relation is used. 

\begin{equation}
	g^2 = \frac{\phi}{p},
\end{equation}
where $1/p$ is the proportionality. We substituted $\sqrt{\phi/p}$ for $g$ in the ABCD matrix to calculate the black line in Supplementary Fig.~\ref{fig:lensing_effect}. The black line is the best fit of the experimental data with the fitting parameters of $d =$ 19.5~mm, $p =$ 300~rad$\cdot$mm$^2$, and $z_0$ = 1.8~mm, where $z_0$ is the distance between the input beam waist and the center of the lens medium. Thus, the experimental data of the output probe beam width as a function of the phase shift can be well described by the effect of the GRIN lens made of the atom cloud.

The DDI among the Rydberg polaritons produced a phase shift at $\delta = 0$ and red-shifted the frequency of $\phi = 0$. To avoid the lensing effect of the atoms as well as to observe the thermalization effect due to the DDI at the high OD, we searched for the two-photon detuning, $\delta_0$, corresponding to $\phi = 0$ for each DDI strength or each value of $\Omega_p^2$. The procedure of searching for $\delta_0$ and then taking the image has been described in \ref{sec:thermalization}. 
 


\section{Relation between the transverse momentum distribution of photons and the intensity profile of an EMCCD image}
\label{sec:derive_TMD}

\MAIN{In the main text}, the transverse momentum distributions in \MAIN{Figs.~5(d)-5(f)} are derived from the images in \MAIN{Figs.5(a)-5(c)}. We measured the size of the probe beam at the far field, while the initial size of the probe beam right after leaving the atom cloud was negligible. For example, the beam size in \MAIN{Fig.~5(c) of the main text} is 1.6 mm and its corresponding beam size at the atom cloud is merely 44~$\mu$m. The idea of the measurement here is similar to the time-of-flight image of cold atoms. As long as the initial size of the atom cloud is negligible as compared with the size of the atom cloud image taken right after a given flight time, the spatial distribution of the image determines the velocity distribution of the atoms. Similarly, we can directly determine the distribution of the transverse momentum of the output photons from the far-field image of the beam size. According to the setup in \MAIN{Fig.~3 of the main text}, the position, $x$ or $y$, on the EMCCD image relates to the emission angle from atom cloud, $\theta$, of photons at the atom cloud as the following;
\begin{equation}
    x=\frac{(s-f)(f-d)+f^2}{f} \theta\equiv \zeta \theta,
    \label{eq:xonCCD}
\end{equation}
where $f$ is the focal length of the lens L3, $s$ is the distance between the atom cloud and lens, and $d$ is the distance between the lens and image plane. Since the maximum value of $\theta$ in the study was less than 0.01 rad., we employed the paraxial approximation in the derivation of the above equation. The transverse momentums, $k_x$ or $k_y$, of photons at the output of the atom cloud is given by
\begin{equation}
    k_x=\frac{2\pi}{\lambda} \theta=\frac{2\pi}{\lambda}\frac{x}{\zeta},
    \label{eq:momentumk}
\end{equation}
where $\lambda$ is the wavelength of the photons and $\zeta\approx$ 204 mm according to \MAIN{Fig.~3 of the main text}. We utilized the above formula to construct the transverse momentum distribution of photons at the atom cloud from the intensity profile of an EMCCD image. \MAIN{Figure~5(a) of the main text} shows the EMCCD image of the probe field without the presence of the atoms. Since the Gaussian beam of the probe field was focused to the center of the atom cloud with the e$^{-1}$ full width of 39~$\mu$m, we can predict the transverse momentum distribution of its photons. The prediction is consistent with the result shown in \MAIN{Fig.~5(d) of the main text}.

